\documentclass[twocolumn]{aastex631}
\pdfoutput=1 
\usepackage{amsmath,amstext}
\usepackage[T1]{fontenc}
\usepackage[figure,figure*]{hypcap}
\usepackage{mwe}
\usepackage{lipsum}
\usepackage{graphicx} 
\usepackage{multirow}
\usepackage{xcolor,colortbl}
\usepackage{xspace}
\usepackage{csvsimple}
\usepackage{afterpage}

\newcommand{\twelveCO}{$^{12}$CO}

\newcommand{\XCO}{$X_\text{CO}$\xspace}

\newcommand{\alphavir}{$\alpha_\text{vir}$\xspace}

\shorttitle{ALMA-LEGUS I: The Influence of Galaxy Morphology}
\shortauthors{Finn et al.}

\begin{document}

\title{ALMA-LEGUS I: The Influence of Galaxy Morphology on Molecular Cloud Properties}

\author[0000-0001-9338-2594]{Molly K. Finn} %
\affiliation{Department of Astronomy, University of Virginia, Charlottesville, VA 22904, USA}

\author[0000-0001-8348-2671]{Kelsey E. Johnson} %
\affiliation{Department of Astronomy, University of Virginia, Charlottesville, VA 22904, USA}

\author[0000-0002-4663-6827]{Remy Indebetouw} %
\affiliation{Department of Astronomy, University of Virginia, Charlottesville, VA 22904, USA}
\affiliation{National Radio Astronomy Observatory, 520 Edgemont Road, Charlottesville, VA 22903, USA}

\author[0000-0002-7408-7589]{Allison H. Costa} %
\affiliation{National Radio Astronomy Observatory, 520 Edgemont Road, Charlottesville, VA 22903, USA}

\author{Angela Adamo}
\affiliation{Department of Astronomy, The Oskar Klein Centre, Stockholm University, SE-106 91 Stockholm, Sweden}

\author[0000-0003-4137-882X]{Alessandra Aloisi}
\affiliation{Space Telescope Science Institute, 3700 San Martin Drive, Baltimore, MD 21218, USA}

\author[0000-0002-7845-8498]{Lauren Bittle}
\affiliation{Independent Researcher}

\author[0000-0002-5189-8004]{Daniela Calzetti}
\affiliation{Department of Astronomy, University of Massachusetts Amherst, 710 North Pleasant Street, Amherst, MA 01003, USA}

\author[0000-0002-5782-9093]{Daniel A. Dale}
\affiliation{Department of Physics \& Astronomy, University of Wyoming, Laramie, WY 82071}

\author[0000-0002-4578-297X]{Clare L. Dobbs}
\affiliation{School of Physics and Astronomy, University of Exeter, Stocker Road, Exeter, EX4 4QL, UK}

\author[0000-0002-3106-7676]{Jennifer Donovan Meyer}
\affiliation{National Radio Astronomy Observatory, 520 Edgemont Road, Charlottesville, VA 22903, USA}

\author[0000-0002-1723-6330]{Bruce G. Elmegreen}
\affiliation{IBM Research Division, T. J. Watson Research Center, 1101 Kitchawan Road, Yorktown Heights, NY 10598, USA}

\author[0000-0002-1392-3520]{Debra M. Elmegreen}
\affiliation{Department of Physics and Astronomy, Vassar College, Poughkeepsie, NY 12604, USA}

\author[0000-0001-6676-3842]{Michele Fumagalli}
\affiliation{Dipartimento di Fisica G. Occhialini, Universit\`a degli Studi di Milano Bicocca, Piazza della Scienza 3, 20126 Milano, Italy}
\affiliation{INAF – Osservatorio Astronomico di Trieste, via G. B. Tiepolo 11, I-34143 Trieste, Italy}

\author[0000-0001-8608-0408]{J. S. Gallagher}
\affiliation{Department of Astronomy, University of Wisconsin-Madison\\ 475 North Charter St., Madison, WI 53706}
\altaffiliation{Department of Physics and Astronomy, Macalester College, 1600 Grand Ave. , St. Paul, MN 55105}

\author[0000-0002-3247-5321]{Kathryn Grasha}
\affiliation{Research School of Astronomy and Astrophysics, Australian National University, Cotter Rd., Weston ACT 2612, Australia} 
\affiliation{ARC Centre of Excellence for Astrophysics in 3D (ASTRO-3D), Canberra ACT 2600, Australia}   
\affiliation{Visiting Fellow, Harvard-Smithsonian Center for Astrophysics, 60 Garden Street, Cambridge, MA 02138, USA}

\author[0000-0002-1891-3794]{Eva K. Grebel}
\affiliation{Astronomisches Rechen-Institut, Zentrum f\"ur Astronomie der Universit\"at Heidelberg, M\"onchhofstr.\ 12--14, 69120 Heidelberg, Germany}

\author[0000-0001-5448-1821]{Robert C. Kennicutt}
\affiliation{Steward Observatory, University of Arizona, Tucson, AZ 85719, USA}
\affiliation{George P. and Cynthia W. Mitchell Institute for Fundamental Physics and Astronomy, Texas A\&M University, College Station, TX 77845, USA}

\author[0000-0003-3893-854X]{Mark R. Krumholz}
\affiliation{Research School of Astronomy and Astrophysics, Australian National University, Cotter Rd., Weston ACT 2612, Australia}
\affiliation{ARC Centre of Excellence for Astrophysics in 3D (ASTRO-3D), Canberra ACT 2600, Australia}

\author[0000-0002-2278-9407]{Janice C. Lee}
\affiliation{Space Telescope Science Institute, 3700 San Martin Drive, Baltimore, MD 21218, USA}

\author[0000-0003-1427-2456]{Matteo Messa}
\affiliation{Observatoire de Gen\'eve, Universit\'e de Gen\`eve, Versoix, Switzerland}
\affiliation{The Oskar Klein Centre, Department of Astronomy, Stockholm University, AlbaNova, SE-10691 Stockholm, Sweden}

\author[0000-0001-7069-4026]{Preethi Nair}
\affiliation{Department of Physics and Astronomy, The University of Alabama, Tuscaloosa, AL 35487, USA}

\author[0000-0003-2954-7643]{Elena Sabbi}
\affiliation{Space Telescope Science Institute, 3700 San Martin Drive, Baltimore, MD 21218, USA}

\author[0000-0002-0806-168X]{Linda J. Smith}
\affiliation{Space Telescope Science Institute, 3700 San Martin Drive, Baltimore, MD 21218, USA}

\author[0000-0002-8528-7340]{David A. Thilker}
\affiliation{Department of Physics and Astronomy, The Johns Hopkins University, Baltimore, MD, 21218 USA}

\author{Bradley C. Whitmore}
\affiliation{Space Telescope Science Institute, 3700 San Martin Drive, Baltimore, MD 21218, USA}

\author[0000-0001-8289-3428]{Aida Wofford}
\affiliation{Instituto de Astronom\'ia, Universidad Nacional Aut\'onoma de M\'exico, Unidad Acad\'emica en Ensenada, Km 103 Carr. Tijuana$-$Ensenada, Ensenada, B.C., C.P. 22860, M\'exico}

\begin{abstract}

We present a comparative study of the molecular gas in two galaxies from the LEGUS sample: barred spiral NGC~1313 and flocculent spiral NGC~7793. These two galaxies have similar masses, metallicities, and star formation rates, but NGC~1313 is forming significantly more massive star clusters than NGC~7793, especially young massive clusters ($<10$ Myr, $>10^4$ M$_\odot$). 
Using ALMA CO(2-1) observations of the two galaxies {with the same sensitivities and resolutions of 13~pc}, we directly compare the molecular gas in these two similar galaxies to determine the physical conditions responsible for their large disparity in cluster formation. 
{By fitting size-linewidth relations for the clouds in each galaxy, we find that NGC~1313 has a higher intercept than NGC~7793, implying that its clouds have higher kinetic energies at a given size scale. NGC~1313 also has more clouds near virial equilibrium than NGC~7793, which may be connected to its higher rate of massive cluster formation. However, these virially bound clouds do not show a stronger correlation with young clusters than that of the general cloud population. }
We find surprisingly small differences between the {distributions of} molecular cloud populations in the two galaxies, though the largest of those differences are that NGC~1313 has higher surface densities and lower free-fall times.

\end{abstract}


\section{Introduction} \label{sec:intro}

\begin{figure*}
    \centering
    \includegraphics[width=0.47\textwidth]{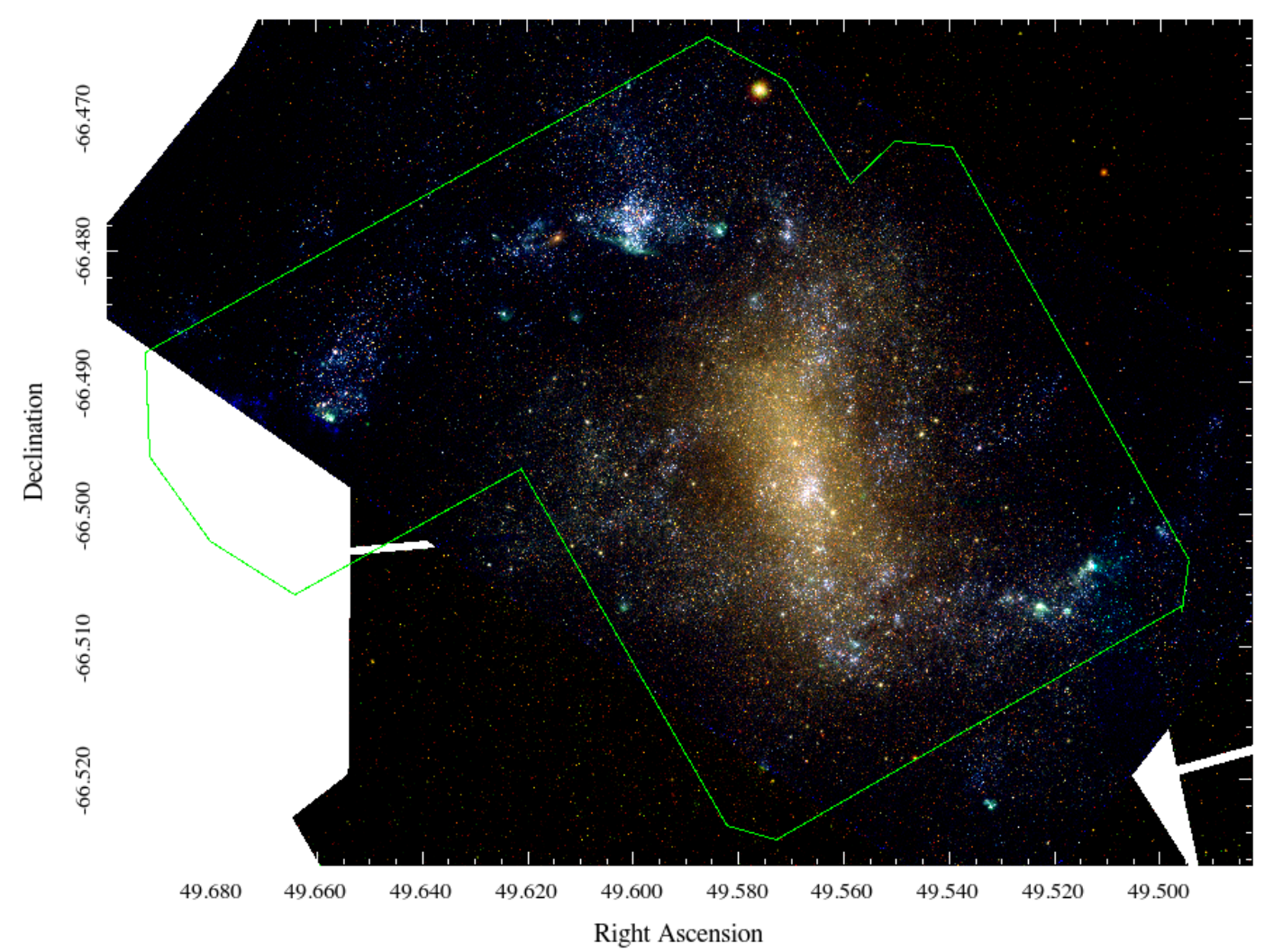}
    \includegraphics[width=0.46\textwidth]{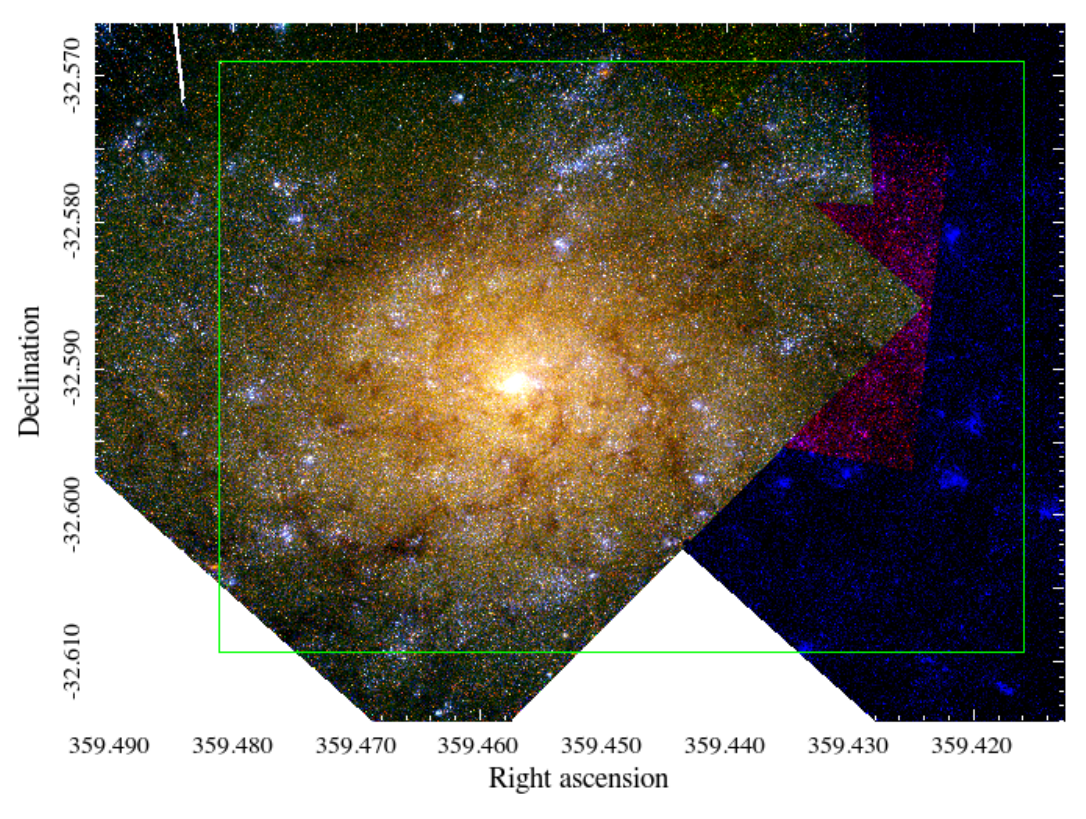}
    \includegraphics[width=0.45\textwidth]{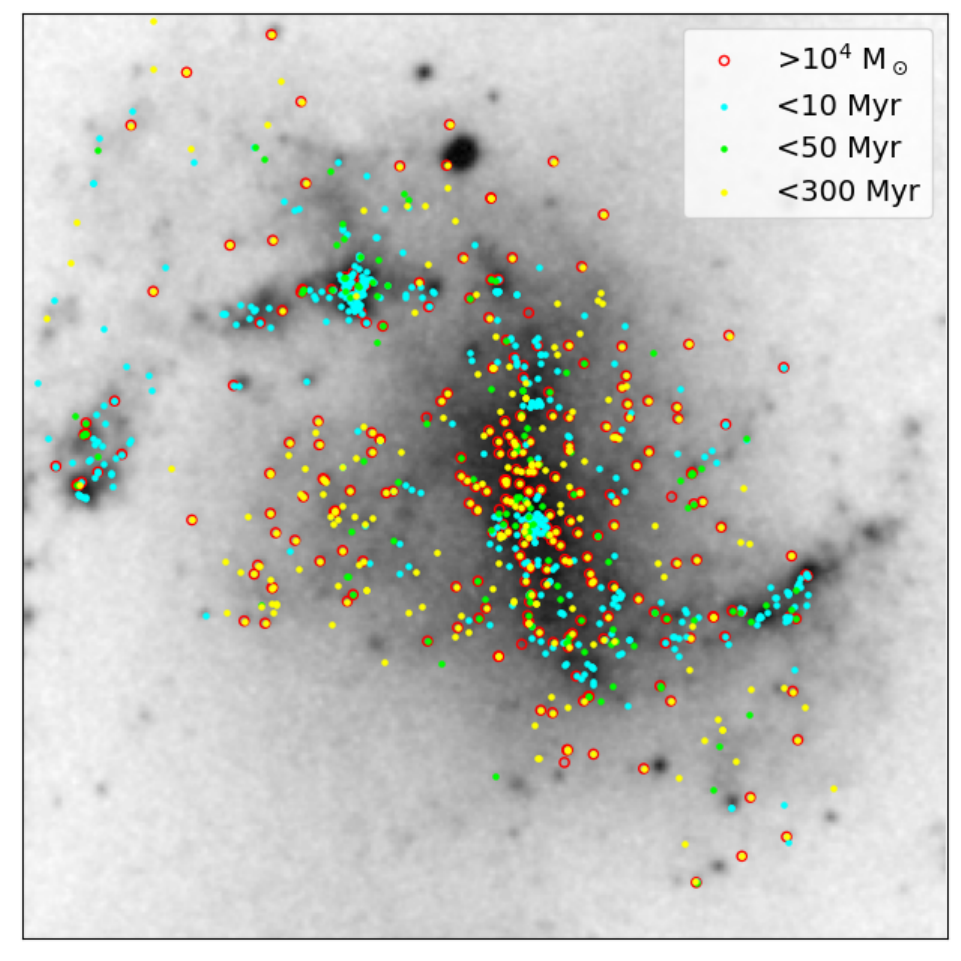}
    \includegraphics[width=0.45\textwidth]{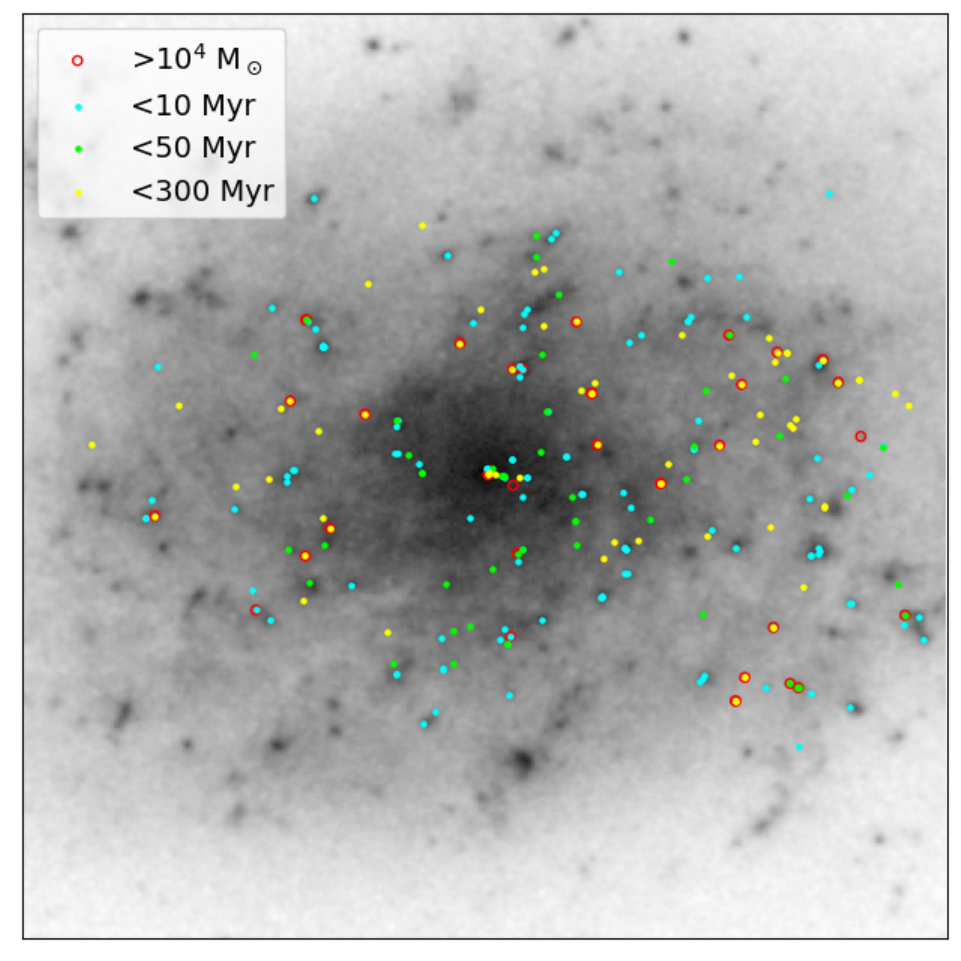}
    \caption{\emph{Top:} Hubble Space Telescope images of NGC~1313 (left) and NGC~7793 (right) using filters F814W (red), F555W (green), and F336W (blue) from the LEGUS survey. Overplotted are the ALMA CO(2-1) observation footprints. \emph{Bottom: } Positions of clusters identified in the LEGUS catalogs (methodology described by \cite{Adamo17}), colored by age with clusters younger than 10 Myr in cyan, 10-50 Myr in green, and 50-300 Myr in yellow, plotted overtop the DSS2-red images. Clusters that are more massive than $10^4 M_\odot$ are outlined in red. NGC~1313 has significantly more star clusters overall than NGC~7793, even after accounting for their differing SFRs, and especially has more red-outlined massive star clusters, both by number and by fraction of the total cluster mass. }
    \label{fig:RGB images}
\end{figure*}

At low redshifts, the majority of star formation takes place in spiral galaxies \citep{Brinchmann04}. This means that to understand star formation in the local universe, we need to understand the influence of spiral structure on the physical conditions of the molecular gas and the processes by which that gas is converted into stars. 

{There have been several studies that have examined the molecular gas properties traced by CO in nearby spiral galaxies. Some examples of these studies include PAWS (PdBI Arcsecond Whirlpool Survey), which mapped M51 in CO(1-0) at 40~pc resolution \citep{Schinnerer13}, CANON (CArma and NObeyama Nearby galaxies), which mapped the inner disks of nearby spiral galaxies in CO(1-0) and enabled a focused study of a subsample at 62-78~pc resolution \citep{JDM13}, and the barred spiral galaxy M83 was mapped in CO(1-0) at 40~pc resolution \citep{Koda23}. These three studies focused primarily on galaxies with strong spiral arms rather than flocculent spirals.} 

{PHANGS-ALMA (Physics at High Angular resolution in Nearby Galaxies) expanded the type of galaxies included, mapping 90 galaxies in CO(2-1) at $\sim100$~pc resolution \citep{Leroy21}, including galaxies with a wide range of morphologies. Comparing the galaxies in their sample by morphology type, \cite{Stuber23} found that the flocculent spirals tended to have lower {total stellar} masses and star formation rates than galaxies with stronger spiral patterns. 
Meanwhile \cite{Hart17} found that in a sample of spiral galaxies from the Sloan Digital Sky Survey, galaxies with two strong spiral arms had similar star formation rates but higher efficiency in converting their gas to stars than galaxies with many flocculent spiral arms. 
While studies have examined how the cloud properties of spiral galaxies vary between the disks, centers, arms, and interarm regions \citep[e.g.][]{Koda09,Rosolowsky08,Colombo14,Sun18PHANGS,Sun20,Rosolowsky21,Querejeta21,Koda23}, a more detailed examination of how the cloud properties vary between galaxies of different morphologies is not yet available. }

\begin{table*}
    \centering
    \caption{Properties of NGC~1313 and NGC~7793}
    \begin{tabular}{l | ccccccc}
    \hline
    \hline
        Galaxy & Distance$^a$ & Metallicity$^b$ & SFR$^c$ & M$_*^d$ & M$_\text{HI}^{e}$ & {{M$_\text{mol}^f$}} & 
        {M$_\text{cluster}^g$ }\\
         & (Mpc) & ($12+\log(\text{O/H})$) & (M$_\odot$/yr) & ($\times10^9$ M$_\odot$) & ($\times10^9$ M$_\odot$) & ($\times10^6$ M$_\odot$) & 
         ($\times10^6$ M$_\odot$) \\
        \hline
        NGC~1313 & 4.6 & 8.4  & 1.15 & 2.6 & 2.1  & 7.6$^{f}$ 
        & 17.3$^{g}$ \\
        NGC~7793 & 3.7 & 8.52 & 0.52 & 3.2 & 0.78 & 14.0$^{f}$
        & 6.4$^{g}$  \\
        \hline
    \end{tabular}
    \label{tab:intro comparison}
    \\
    \raggedright
    {Notes:} 
    $^a$ \cite{Qing15}, \cite{Gao16} for NGC~1313; \cite{RadburnSmith11}, \cite{Sabbi18} for NGC~7793, \\
    $^b$ \cite{WalshRoy97} and \cite{Stanghellini15}, \\
    $^c$ Calculated by \cite{Calzetti15} using GALEX far-UV images, dust corrected with methods of \cite{Lee09}, \\
    $^d$ Calculated by \cite{Calzetti15} using extinction-corrected $B$-band luminosities and methods of \cite{Bothwell09}, \\
    $^e$ Calculated by \cite{Calzetti15} using HI observations from \cite{Koribalski04}, \\
    {$^f$ Sum of all CO(2-1) emission, multiplied by an average \XCO and factor of 1.4 for each galaxy as discussed in Section\,\ref{sec:properties}, does not cover full galaxies due to smaller observational footprint and is likely missing flux because 7m and TP data are not included here, }\\
    $^g$ Total mass of star clusters in LEGUS catalogs, based on methods of \cite{Adamo17}, does not cover full galaxies.
\end{table*}

{To better understand the role of galaxy morphology on molecular gas conditions on the scale of individual clouds and how it relates to star formation,} we present a comparative study of the molecular gas in two spiral galaxies: NGC~1313 and NGC~7793 (Figure\,\ref{fig:RGB images}). These two galaxies are included in the Legacy ExtraGalactic UV Survey \citep[LEGUS;][]{Calzetti15}, a Hubble Space Telescope (HST) Treasury Program that observed 50 nearby ($<12$ Mpc) galaxies. As part of the LEGUS survey, they both have comprehensive catalogs of their young star clusters and their masses and ages using the methodology described in \cite{Adamo17}, which allows us to compare the star-forming properties of each galaxy with the molecular gas at scales of $\sim10$~pc. 

NGC~1313 and NGC~7793 were chosen because they have many similar properties, such as their total stellar mass \citep[$2.6\times10^9$ and $3.2\times10^9$ M$_\odot$;][]{Calzetti15}, their overall metallicities \citep[$12+\log(\text{O/H})=$ 8.4 and 8.52;][]{WalshRoy97,Stanghellini15}, their star formation rates \citep[SFR; 1.15 and 0.52 M$_\odot$/yr {from dust-attenuation-corrected far-UV images};][]{Calzetti15}, and their Hubble type of Sd. They are also both mostly face-on with clear views of the spiral structures in each \citep[inclinations of $40.7\deg$ and $47.4\deg$;][]{Calzetti15} and have similar distances \citep[4.6 and 3.7 Mpc;][]{RadburnSmith11,Qing15,Gao16,Sabbi18}. Their main properties are listed in Table\,\ref{tab:intro comparison}.

Despite these many similarities, the two galaxies have starkly different spiral structures, with NGC~1313 being barred with strong spiral arms and NGC~7793 being a flocculent spiral (see Figure\,\ref{fig:RGB images}). NGC~1313 also appears to be experiencing a minor interaction (see \S\ref{subsec:NGC1313 intro}). Based on the LEGUS cluster catalogs, they also have strikingly different numbers of massive clusters. NGC~1313 has more than six times as many clusters that are more massive than $10^4$ M$_\odot$, despite having a SFR only 2.2 times higher than NGC~7793. 
{NGC~1313 also has 2.7 times more neutral HI gas \citep[$2.1\times10^9$ and $0.78\times10^9$ M$_\odot$;][]{Calzetti15}, but only half as much molecular gas as NGC~7793, as measured in the observational footprint presented here (see Section\,\ref{sec:properties}).} This makes NGC~1313 and NGC~7793 an interesting pair of galaxies to compare to further understand how the spiral structure of a galaxy influences the molecular gas properties and the types of star clusters that are formed. 

\subsection{NGC~1313} \label{subsec:NGC1313 intro}

NGC~1313 has an irregular morphology with a strong bar and asymmetric spiral arms, which has historically been compared to the Large Magellanic Cloud \citep{deVaucouleurs63}. Its irregular morphology \citep{SandageBrucato79} and observations of the HI showing a loop of gas around the galaxy and a disturbance in its velocity field in the southwest of the galaxy \citep{Peters94} indicate that it is interacting with a satellite galaxy. When measuring the galaxy's star formation history, \cite{Larsen07} found an increase in recent star formation in the southwest of the galaxy, potentially caused by this interaction. This was further confirmed by \cite{SilvaVillaLarsen12}, who found that rather than undergoing a starburst across the whole galaxy, there was only an increase in star formation in the southwestern field. They find that the regional starburst occurred $\sim100$ Myr ago. 

{\cite{WalshRoy97} found that NGC~1313 is one of the most massive galaxies that has no gas-phase metallicity gradient across its disk. However, the more recent study \cite{Hernandez22}} shows evidence for a metallicity gradient based on the chemical abundances of young clusters. They also find a constant star formation rate across the disk of NGC~1313 with the exception of the burst in the southwest.  This southwestern region where the localized starburst is observed is just outside of the observation footprint for this study. A difference in stellar metallicity tracing a separate population was found throughout the disk by \cite{Tikhonov16}, which they attribute to a past merger with a low-mass dwarf satellite that did not gravitationally distort the arms or central region of NGC~1313.

The star cluster population in NGC~1313 has been characterized using the LEGUS cluster catalogs by \cite{Grasha17b}, and the sizes of these clusters were further characterized by \cite{Ryon17}, who found that the clusters are undergoing relaxation but appear to be gravitationally bound. \cite{Hannon19} studied the morphology of the H$\alpha$ around these clusters and report clearing times for the gas of $\lesssim1$ Myr. \cite{Messa21} added to the LEGUS catalogs by searching for embedded clusters identified with Pa$\beta$ and in the near-infrared and finding that up to 60\% of the star clusters are not accounted for in the UV-optical catalogs {and that gas-clearing timescales are closer to 3-4~Myr}.

\subsection{NGC~7793} \label{subsec:NGC7793 intro}

NGC~7793 is a member of the Sculptor group of galaxies and has been well-studied due to its proximity and high Galactic latitude. It has no bar, a small bulge with a nuclear star cluster \citep{Kacharov18}, and a mostly uniform distribution of small, loose spiral arms with little coherent structure \citep{Elmegreen14}, to the point that it has been called an ``extreme'' flocculent spiral \citep{Elmegreen84}. It has nearby dwarf companions, and the HI disk is warped, suggesting some level of interaction, though they see no sign of tidal effects \citep{Koribalski18}. It has drawn attention for its unusual discontinuous and positive metallicity gradient in the far outer disk \citep{Vlajic11} and its declining rotation curve in the outer edges of the disk \citep{CarignanPuch90, Dicaire08}, although the latter could be due to a line-of-sight warp in the disk \citep{Bacchini19}. Studies of its stellar population have also shown that it has a break in the disk {where the surface brightness departs from simple exponential decrease}, likely caused by radial migration of the stars \citep{RadburnSmith12}.  

The star formation history of NGC~7793 has been studied by \cite{Sacchi19} and they found an inside-out growth of the galaxy, with the outer regions undergoing a greater recent increase in star formation than the inner region. Inside-out growth was also seen in young clumps identified in UV light by \cite{Mondal21}, who also suggest that NGC~7793 is experiencing a recent increase in its star formation rate. Its LEGUS-identified star clusters were included in the same studies as NGC~1313 by \cite{Grasha17b}, \cite{Hannon19}, and \cite{BrownGnedin21} to characterize their populations, cluster boundedness, and gas clearing timescales. The clusters and molecular gas were further studied by \cite{Grasha18}, finding that the younger clusters are more spatially correlated with molecular clouds and that the hierarchical clustering of the clouds is shared by the young clusters. \cite{Muraoka16} has also mapped the CO(3-2) line in NGC~7793, finding that its emission is well-correlated with infrared tracers of star formation.

\hfill \\

To robustly compare the properties of the molecular clouds in this pair of galaxies, we have observed both in CO(2-1) with carefully matched spatial resolutions of 13~pc and surface brightness sensitivities of $\sim0.2$ K. These observations allow us to make a direct comparison of clouds in a galaxy with strong spiral arms and in a flocculent spiral without concern for resolution or sensitivity effects. This paper is organized as follows: 
We present the observations used in this analysis in Section\,\ref{sec:obs}, and the LEGUS cluster catalogs of the two galaxies in Section\,\ref{sec: cluster catalogs}. We then discuss the decomposition of the CO(2-1) emission into substructures in Section\,\ref{sec:dendrogram} and the calculation of their properties in Section\,\ref{sec:properties}. We examine the size-linewidth relations of the clouds in each galaxy in Section\,\ref{sec:SL plots}, and the virialization of the clouds in Section\,\ref{sec: virial plot}. We compare the distributions of all the measured and derived properties in Section\,\ref{sec:property comparison}. {In Section\,\ref{sec:spatial distributions}, we examine how the clouds and their properties are spatially correlated with those of the clusters. }
We discuss our findings in Section\,\ref{sec:discussion} and summarize our conclusions in Section\,\ref{sec:conclusions}.

\section{Observations} \label{sec:obs}

\begin{figure*}
    \centering
    \includegraphics[width=0.93\textwidth]{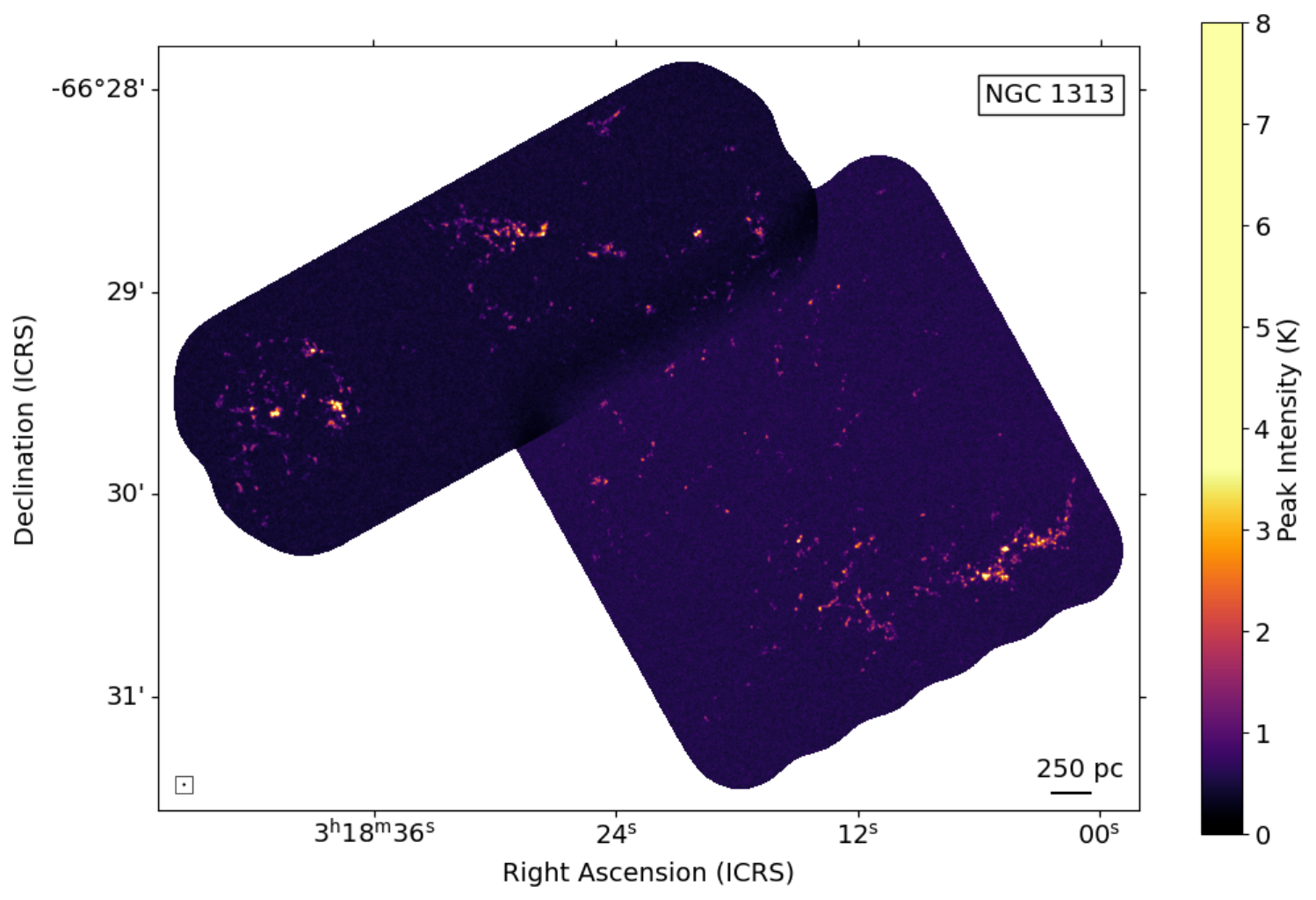}
    \includegraphics[width=0.93\textwidth]{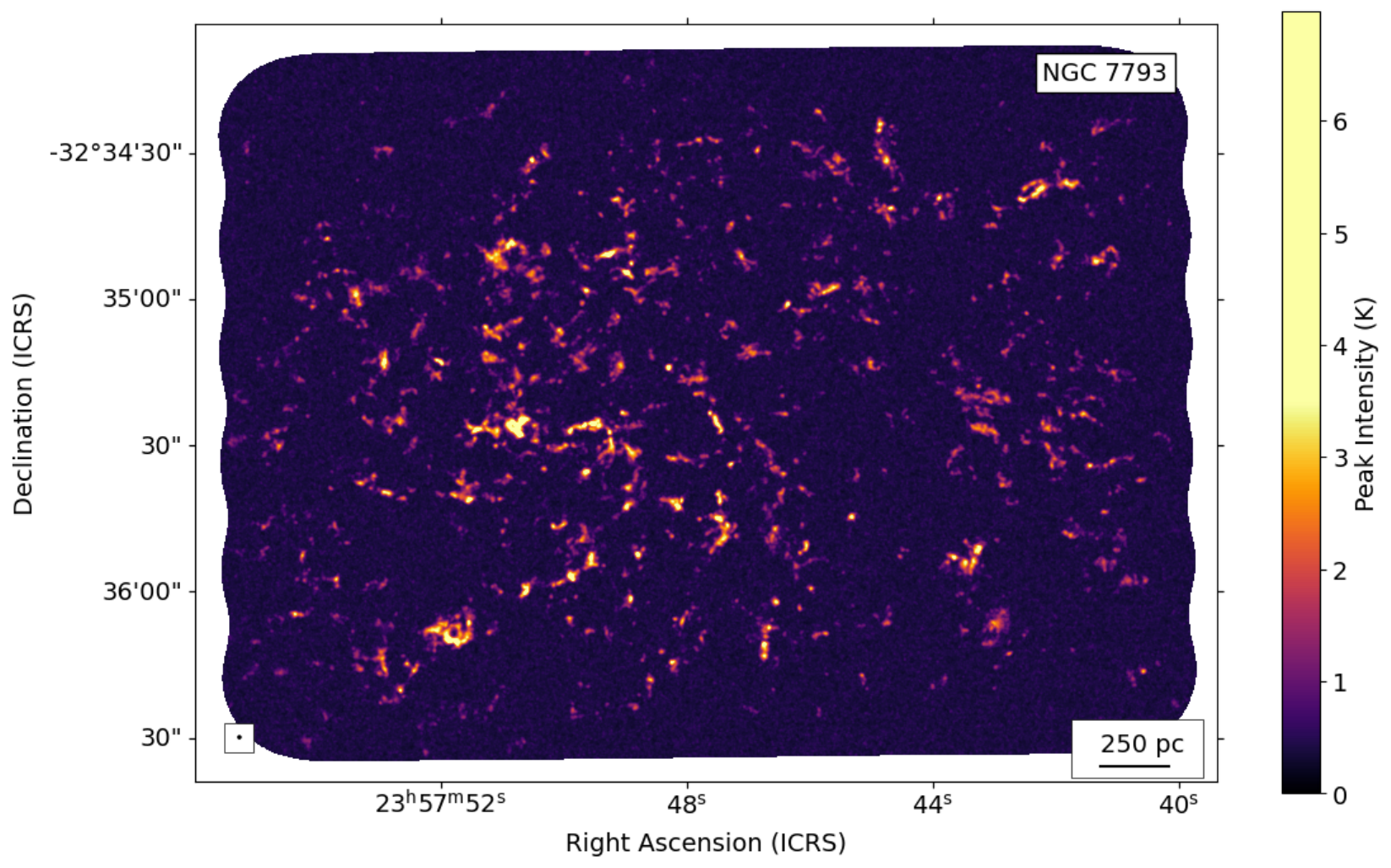}
    \caption{CO(2-1) peak intensity maps of NGC~1313 (top) and  NGC~7793 (bottom), with beam sizes shown in the bottom left corners.}
    \label{fig:maps}
\end{figure*}

\subsection{NGC~1313 CO(2-1)} \label{subsec: 1313 obs}

NGC~1313 was observed by ALMA in Band 6 (project code 2015.1.00782.S; PI: K.~E. Johnson) with the {12m array}. It was originally observed in Cycle 4 in 2016 but did not meet sensitivity requirements and so was observed again in Cycle 5 in 2018. The galaxy is split into two mosaics, one covering the northern arm of the galaxy, and one covering the central region and southern arm. The northern arm mosaic consists of 68 pointings with 1.9 minutes of integration per pointing  with a spacing of 12.9\arcsec\ for a total integration time of 2.2 hours. The central mosaic consists of 104 pointings with 2.2 minutes of integration per pointing for a total integration time of 3.75 hours, also with a spacing of 12.9\arcsec. {The largest angular scale recovered with these observations is 7.4\arcsec, which corresponds to a spatial scale of 150~pc, much larger than any of the clouds measured in this analysis.}

The data were calibrated with the ALMA data pipeline version Pipeline-CASA54-P1-B \citep{Hunter23} in CASA 5.4.0-68 using  J2258-2758 for bandpass and amplitude calibration and J2353-3037 for phase calibration. It was imaged in CASA 6.1.1.15 with a robust parameter of 0.5, resulting in a synthesized beam of 0.579\arcsec$\times$0.486\arcsec (12.8$\times$11.2~pc at a distance of 4.6 Mpc). 
In Figure\,\ref{fig:maps}, we show the image data after smoothing to a circular beam size of 13 pc to match the observations of NGC~7793, which corresponds to 0.583\arcsec for NGC~7793. A summary of the resulting image is shown in Table\,\ref{tab:observations}.

\subsection{NGC~7793 CO(2-1)} \label{subsec: 7793 obs}

NGC~7793 was observed by ALMA in Band 6 during Cycle 4 in 2016 (project code 2015.1.00782.S; PI: K.~E. Johnson) with the {12m array}. The observations were a mosaic of the central 180\arcsec$\times$114\arcsec\ (3$\times$2 kpc) of NGC 7793. The total integration time is 3 hours with the 12-m array with 148 pointings at 1.2 minutes of integration per pointing and a mosaic spacing of 12.9\arcsec\ between pointings. {The largest angular scale recovered with these observations is 7.0\arcsec, which corresponds to a spatial scale of 125~pc, much larger than any of the clouds measured in this analysis.}

The data were calibrated with the ALMA data pipeline version 2020.1.0-40 \citep{Hunter23} in CASA 6.1.1.15 using J0519-4546 and J0334-4008 for bandpass and amplitude calibration and J0303-6211 for phase calibration. It was imaged using a Briggs weighting with a robust parameter of 2.0, resulting in a synthesized beam size of 0.686\arcsec$\times$0.595\arcsec (12.3$\times$10.7~pc at a distance of 3.7 Mpc). 
In Figure\,\ref{fig:maps}, we show the image data after smoothing to a circular beam size of 13 pc to match the observations of NGC~1313, which corresponds to 0.724\arcsec for NGC~7793. A summary of the resulting image is shown in Table\,\ref{tab:observations}.
These data also appeared in \cite{Grasha18}.

\begin{table}
    \centering
    \caption{ALMA \twelveCO(2-1) Observations}
    \begin{tabular}{ccccc}
    \hline
    \hline
        Galaxy & Beam & Beam & rms & Velocity \\
         & (arcsec) & (pc) & (K) & Resolution (km/s)\\
         \hline
        NGC~1313 & 0.58 & 13 & 0.15 & 1.33 \\
        NGC~7793 & 0.72 & 13 & 0.2  & 1.33 \\
        \hline
    \end{tabular}
    
    \label{tab:observations}
\end{table}

\section{Cluster Catalogs} \label{sec: cluster catalogs}

In addition to the CO(2-1) data, we use catalogs of star clusters and their SED-fitted properties from the LEGUS collaboration, which used the methodology described in \cite{Adamo17}. In this work, we use their catalog that uses {the Milky Way extinction law from \cite{Cardelli89}}, averaged aperture correction method, and Padova stellar evolutionary tracks implemented by Yggdrasil \citep{Zackrisson11}. We include all objects identified by LEGUS as cluster candidates, which required that they were brighter than M$_{V}$ of -6, detected above $3\sigma$ in the UBVI set of filters, and had a visual classification of 1 or 2 (compact) or class 3 (multiply-peaked). 
\cite{Adamo17} adopt a 90\% completeness limit of 5000~M$_\odot$ for clusters with ages up to 200 Myr in NGC~602, which has a distance of 10~Mpc. 
However, at the distances of NGC~1313 and NGC~7793, we expect this limit to be much lower, closer to 1000 M$_\odot$ \citep{Grasha18}. We represent this 1000 M$_\odot$ limit as a vertical line in the mass distributions shown in Figure\,\ref{fig:global cluster distributions}. 

Several studies have pointed out that a degeneracy in age and reddening during the fitting results in many old globular clusters incorrectly being assigned much younger ages \citep{Turner21,Hannon19,Whitmore23}. In particular, \cite{Whitmore23} find that the majority of these objects with incorrectly fitted ages fall above a line extending from (6, 1) to (9, 0.1) in a plot of E(B-V) versus log(Age). They also find that these incorrect ages primarily affect objects on the high mass end ($M>10^{4.5}M_\odot$). To account for these incorrect ages without biasing our sample against high-mass objects, we remove the clusters that fall above the E(B-V)-log(Age) line from \cite{Whitmore23} only when comparing age measurements between the two galaxies. This affects 76 clusters in NGC~1313 and 29 clusters in NGC~7793 {(6\% of the identified clusters in both galaxies)}. Carefully fitting these objects with the degeneracy in mind would likely result in many of them having ages older than 1~Gyr. We also note that there are few enough of these objects that keeping or removing them does not substantially alter our results.

\cite{OrozcoDuarte22} compared the output of the LEGUS cluster catalog for NGC~7793 with results obtained using synthetic photometry and a stochastically sampled IMF and found masses that were on average 0.11~dex larger and ages that were 1~Myr younger compared to the \cite{Adamo17} method. However, we use the original LEGUS catalogs still to maintain a consistent method between the two galaxies.

{\cite{Messa21} also identify a significant population of embedded clusters in NGC~1313. However, since we do not also have embedded clusters identified in NGC~7793 to compare, we do not include those clusters in this work. }

The cluster catalogs for both galaxies were separated into east and west portions of the map. To combine these, we identified matching pairs of clusters in the overlap region. We then used the cluster properties from the map that had the better fit, based on their reported quality of fit  Q parameter. We identified 67 overlapping clusters in NGC~1313 and 47 in NGC~7793. Each of these clumps has a fitted mass, age, and extinction from the catalog.


\subsection{Cluster Counts}

\begin{table*}
    \centering
    \caption{Number of cloud structures and clusters in each galaxy}
    \begin{tabular}{ll | cc}
        \hline
        \hline
        && NGC~1313 & NGC~7793 \\
        \hline
        \multirow{4}{*}{LEGUS Clusters} & All Clusters & 1201 & 467 \\
        & Massive Clusters$^a$ & 333 & 53 \\
        & Young, Massive Clusters$^a,b$ & 37 & 3 \\
         \hline
         \multirow{5}{*}{Molecular Gas Structures} & Trunks$^c$ & 65 & 130 \\
        & Branches$^c$ & 82 & 187 \\
        & Leaves$^c$ & 442 & 761 \\
        & Clumps$^d$ & 531 & 965 \\
        & Massive Clumps$^a,d$ & 137 & 306 \\
        \hline
    \end{tabular} \\
    \raggedright
    {Note:} $^a$The threshold used to define ``massive'' for both clumps and clusters in this table is $M>10^4 M_\odot$. \\ $^b$ The threshold used to define ``young'' for clusters is an age~$<10$~Myr. \\
    {$^c$ Structures identified by \texttt{astrodendro}.} \\
    {$^d$ Structures identified by \texttt{quickclump}.}
    \label{tab:segmentation}
\end{table*}

NGC~1313 has $\sim2.6$ times as many identified star clusters at all masses as NGC~7793, with a total of 1201 clusters compared to 467 for NGC~7793. Considering that the star formation rate in NGC~1313 is $\sim2.2$ times larger than that of NGC~7793, this represents a small excess in cluster formation. We can also account for the fraction of the total SFR included in the footprints of the LEGUS observations used to identify these clusters based on the fraction of the GALEX far-UV light. We note that the SFR traced by UV light is well-matched to the ages of the general cluster population, but may be averaged over too long a timescale to be accurate for the youngest clusters. In NGC~1313, 57\% of the total SFR is included in the LEGUS footprint, and in NGC~7793 53\% is included. Assuming that the cluster populations are well-matched to the SFR within each galaxy, NGC~1313 would have $\sim2110$ clusters and NGC~7793 would have $\sim880$ clusters within the total star forming area of each galaxy. This suggests then that NGC~1313 has $\sim2.4$ times as many clusters than NGC~7793 in the full galaxy, still a slight excess over the difference in total SFR. 

This difference in cluster number is more extreme when we consider only the most massive clusters with $M_* > 10^4 M_\odot$. NGC~1313 has over six times as many massive clusters, having 333 massive clusters identified by LEGUS where NGC~7793 has 53. Correcting for the fraction of star formation traced by LEGUS would still leave NGC~1313 with 5.8 times more massive clusters. Furthermore, 37 of those massive clusters are young (<10 Myr) in NGC~1313, compared to only 3 in NGC~7793. As seen in Figure\,\ref{fig:global cluster distributions}, NGC~7793 does have a couple {of} particularly massive clusters, and so we can also compare the mass of clusters in both {galaxies} above the $M_* > 10^4 M_\odot$ threshold and we find that NGC~1313 has 2.7 times as much mass in massive clusters (2.5 times the mass when including a total SFR correction), which still represents an excess of massive cluster formation in NGC~1313 by this metric. 

{We also note, however, that based on the \cite{Lee09} comparison of the dust-corrected UV-based method of calculating SFR to measuring SFR with H$\alpha$, we estimate that the SFR values reported in \cite{Calzetti15} would have $1\sigma$ uncertainties of approximately $25\%$. \cite{Calzetti15} do not report uncertainties on the SFR for NGC~1313 and NGC~7793 specifically, but if we take this estimated uncertainty value from \cite{Lee09} and propagate the error, we get a ratio of 2.2$\pm$0.8 in the SFR between the two galaxies. In this case, the ratios reported here are not statistically significant when considering error from counting statistics.
}

{Overall,} NGC~1313 not only has more clusters, but also has more massive clusters than NGC~7793, despite the lower molecular gas content of NGC~1313 (as shown in Table\,\ref{tab:intro comparison} and will be discussed in Section\,\ref{sec:properties}). These cluster numbers are shown in Table\,\ref{tab:segmentation}.

\subsection{Cluster Property Distributions}

We further consider how the distributions of cluster masses and ages compare between the two galaxies by looking at histograms, Gaussian kernel density estimations (KDEs) from \texttt{scipy} \citep{scipy}, and cumulative distribution functions (CDFs). Histograms and KDEs are intuitive representations of how the values of a parameter are distributed, but they are also heavily affected by our choice in binning for the histogram and bandwidth for the KDE. We use a scalar estimator bandwidth of 0.5 dex for all KDEs in this section and Section\,\ref{sec:property comparison} for uniformity (Scott's rule would result in bandwidths between 0.2 and 0.5 dex for the various distributions). CDFs are less intuitive for understanding how values of a property are distributed, but they are unaffected by binning and so are the most robust depiction of property distributions. When interpreting CDFs, a line further to the right of the plot indicates a distribution with larger values.

\begin{figure*}
    \centering
    \includegraphics[width=0.45\textwidth]{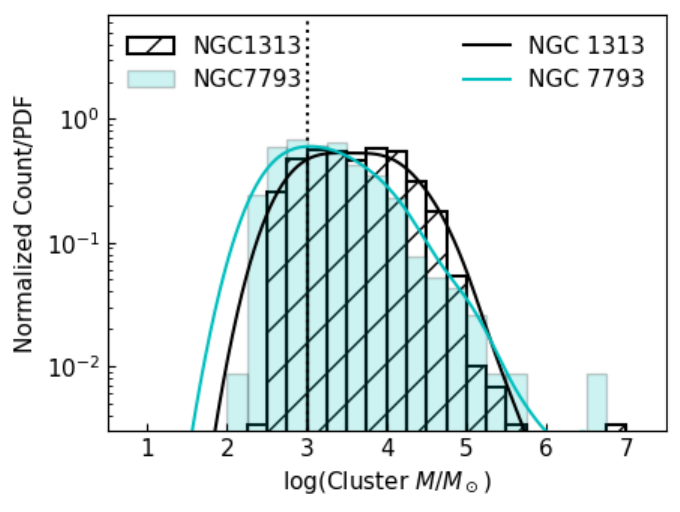}
    \includegraphics[width=0.43\textwidth]{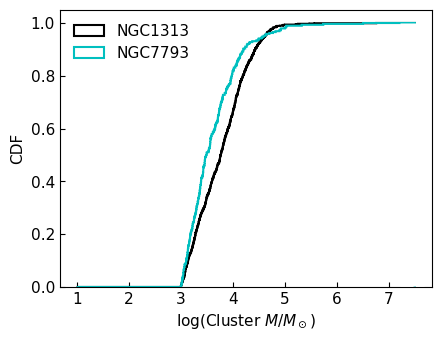}
    \includegraphics[width=0.45\textwidth]{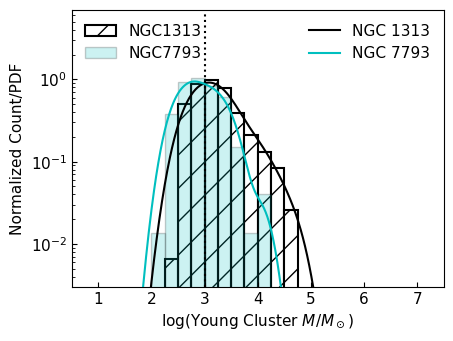}
    \includegraphics[width=0.43\textwidth]{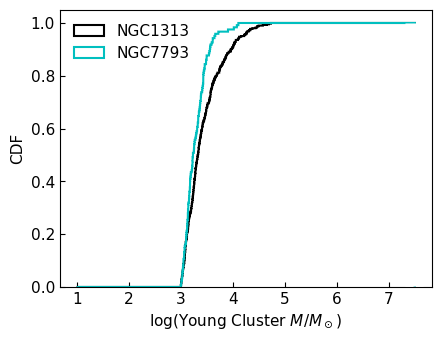}
    \includegraphics[width=0.45\textwidth]{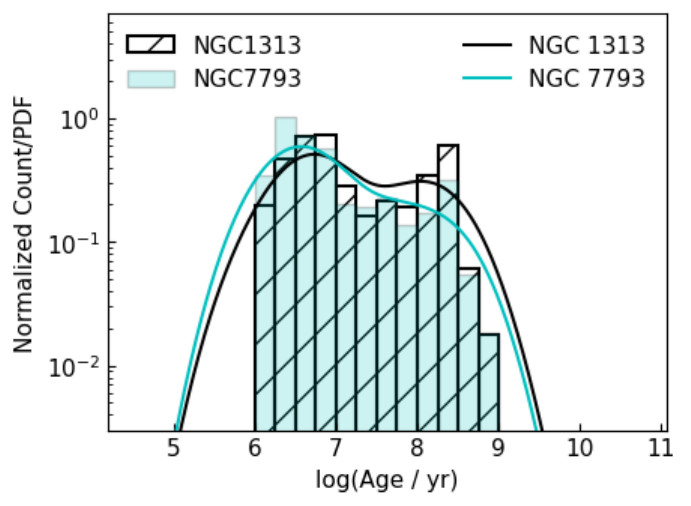}
    \includegraphics[width=0.43\textwidth]{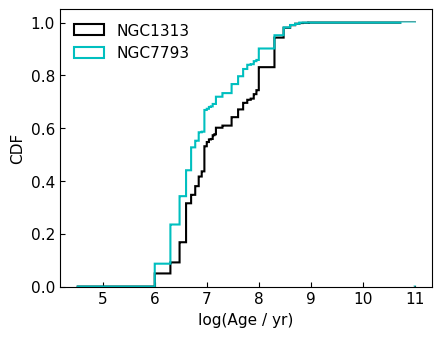}
    \caption{Normalized distributions of the cluster parameters for the two galaxies, using histograms and KDEs (left) and CDFs (right). {The properties shown are the mass distribution for the full cluster population (top), the mass distribution for only the young (<10 Myr) clusters (middle), and the age distribution of all the clusters (bottom). The estimated mass completeness limit of 1000~M$_\odot$ is shown as a vertical line in the mass distributions, and the CDFs of the mass distributions only include clusters above this mass limit.} The age distributions do not include clusters that are likely to have incorrect ages due to the age/reddening degeneracy \citep{Whitmore23}.}
    \label{fig:global cluster distributions}
\end{figure*}

In Figure\,\ref{fig:global cluster distributions}, we show the distributions of the fitted masses {for both the full cluster population and for only the youngest (<10 Myr) clusters, as well as distributions of the fitted ages of the clusters in each galaxy.
We show} histograms and KDEs in the left column and show CDFs in the right column. The cluster masses in NGC~1313 {for both the full population and especially for the youngest clusters are skewed towards higher masses than in NGC~7793, further emphasizing the tendency towards more massive clusters in NGC~1313.} The distribution of cluster ages in the two galaxies appears more similar, though NGC~1313 has slightly older clusters than NGC~7793.

\section{Molecular Gas Structure Decomposition} \label{sec:dendrogram}

To understand the properties of individual molecular gas structures, we use two different methods of emission decomposition to more robustly compare physical conditions between the two galaxies. One method is to use a dendrogram to hierarchically categorize the structures, and the other is to identify non-overlapping clumps. These segmentation methods are described in more detail below. Dendrograms are superior for examining the full spatial scale of molecular clouds in the region because they are able to capture the hierarchical nature of the gas, from large GMCs to smaller knots. However, dendrograms {count emission multiple times} and so they cannot be used for any counting statistics. We therefore use dendrograms in Sections \ref{sec:SL plots} and \ref{sec: virial plot} where we want to understand the full spatial scale of the clouds and multiply-counted emission is allowable. We use the non-overlapping clumps in Section \ref{sec:property comparison}, which focuses on property distributions and so cannot include multiply-counted emission.

\subsection{Dendrogram Segmentation} \label{subsec:dendro}

We use the package \texttt{astrodendro} \citep{Rosolowsky08} to decompose the structures in each galaxy into dendrograms. 
This results in a hierarchical ``tree'' of structures that merge together at lower contour levels. The local maxima are called ``leaves'' and have no further substructure, while the larger merged structures are ``branches'' and ``trunks''. Trunks are not bounded by any other structures. By convention, isolated structures that have no substructure and are also not bounded by any other structure are called leaves instead of trunks. 
We use the input parameters \texttt{min\_value}=3$\sigma$, \texttt{min\_delta}=2.5$\sigma$, and \texttt{min\_npix}=2 beams, where \texttt{min\_value} is the minimum intensity value of an identified emission peak, \texttt{min\_delta} is the minimum intensity separation between structures merging, and \texttt{min\_npix} is the minimum number of voxels in an identified structure. 
The breakdown of each type of dendrogram structure for each galaxy is shown in Table\,\ref{tab:segmentation}.

\subsection{Clump Segmentation} \label{subsec:clump}

We use the algorithm \texttt{quickclump} \citep{Sidorin17} to decompose the emission into clumps that have no overlap. 
This algorithm outputs a similar style of clump decomposition to \texttt{clumpfind} \citep{Williams95}, for example. 
We used the input parameters \texttt{Tcutoff}=4$\sigma$, \texttt{dTleaf}=4$\sigma$, and \texttt{Npixmin}=2 beams (equal to 65 and 70 for NGC~1313 and NGC~7793, respectively). 
\texttt{Tcutoff} is the minimum intensity included in clump assignments, \texttt{dTleaf} is the minimum intensity differences for an emission peak to be considered a separate clump, and \texttt{Npixmin} is the minimum number of voxels in a clump.
The number of clumps identified in each galaxy are shown in Table\,\ref{tab:segmentation}.

\section{Calculating Cloud Properties} \label{sec:properties}

\subsection{Mass}

We calculate the mass of each structure using a CO-to-H$_2$ factor, \XCO, where \XCO = $N_{H_2}$/$W_{CO}$, $N_{H_2}$ is the column density of H$_2$ in cm$^{-2}$ and $W_{CO}$ is the observed brightness of CO in K km s$^{-1}$. We use the $X_\text{CO(2-1)}$ calibration from \cite{Gong20}, Table~3, Equation~3b, which determines the $X_\text{CO(2-1)}$ conversion factor between $N_{H_2}$ and CO(2-1) based on the peak brightness temperature of the clump ($T_\text{peak}$ in K), the beam size ($r_\text{b}$; 13~pc), and the metallicity ($Z$) with the equation

\begin{equation}
    X_{CO(2-1)} = (2.7\times10^{20}) T_\text{peak}^{-1.07+0.37\log{(r_\text{b})}}Z^{-0.5}r_\text{b}^{-0.13}.
    \label{eq:XCO}
\end{equation}

Using this equation, which is calibrated directly to the CO(2-1) line rather than CO(1-0), means that we do not need to assume a ratio of CO(2-1)/CO(1-0), which is known to vary across galaxies \citep{Koda12, Leroy22}.
For NGC~1313, we use an oxygen abundance of $12+\log{\text{(O/H)}}=8.4\pm0.1$ with no radial gradient in the galaxy \citep{WalshRoy97}. NGC~7793 has a measured radial gradient of $12+\log{\text{(O/H)}}=8.572 - 0.054$ dex kpc$^{-1} \times R_\text{gal}$  \citep{Stanghellini15}. Using a solar oxygen abundance of $12+\log{\text{(O/H)}}=8.69$ \citep{Asplund09}, these abundances correspond to metallicities of $Z=0.51Z_\odot$ for NGC~1313 and a range of $Z=(0.56-0.76)Z_\odot$ for NGC~7793. Using the prescription from Eq.\,\ref{eq:XCO} results in values of $X_\text{CO(2-1)}=(0.69-2.56)\times10^{20}$ cm$^{-2}$/(K km s$^{-1}$) for NGC~1313 and \XCO$=(0.64-2.65)\times10^{20}$ cm$^{-2}$/(K km s$^{-1}$) for NGC~7793. 

We also consider the metallicity gradient measured for NGC~7793 by \cite{Grasha22}, which instead finds $12+\log{\text{(O/H)}}=8.945 - 0.083$ dex kpc$^{-1} \times R_\text{gal}$, resulting in metallicities of $Z=(1.14-1.79)Z_\odot$ and values of $X_\text{CO(2-1)}=(0.43-1.83)\times10^{20}$ cm$^{-2}$/(K km s$^{-1}$). We find that the difference in metallicity prescription and resulting masses does not significantly affect any of the results in this paper and so we use the lower metallicities of \cite{Stanghellini15} throughout for NGC~7793. Hereafter, we refer to $X_\text{CO(2-1)}$ as \XCO.

We next calculate the mass of each structure using these calibrated \XCO values to determine the H$_2$ column density, then multiply by the pixel size in cm$^2$ and sum over all pixels to get the mass of H$_2$. We then multiply this M$_{H_2}$ by a factor of 1.4 (which is the mean mass per hydrogen atom and assumes all the hydrogen is molecular) to get the total mass of the gas, $M$. We adopt a 10\% error on the resulting masses due to the standard 10\% ALMA flux calibration uncertainty \citep{Fomalont14}, which we add in quadrature with the error from the measured rms noise (Table\,\ref{tab:observations}). 

We also estimate a total mass of {molecular gas in the observational footprint of each galaxy, using the total CO(2-1) emission multiplied by the respective average \XCO measured for the clumps in each galaxy and multiplying by a factor of 1.4 as mentioned above. We find that NGC~7793 contains nearly twice the molecular mass as NGC~1313, with measured totals of $7.6\times10^6$ M$_\odot$ in NGC~1313 and $1.4\times10^7$ M$_\odot$ in NGC~7793.} {It is important to note} that our observations do not cover the whole galaxy {and we do not include 7m and total power data, so we are likely missing diffuse emission and these values should not be considered the total molecular masses of the galaxies}. The ALMA observations do, however, cover a similar fraction of the galaxy to the LEGUS observations that clusters are derived from. 

{These values of observed molecular gas mass are low compared to those measured by PHANGS-ALMA in \cite{Leroy21}, likely due to the smaller observational footprint and the missing diffuse gas emission since we only use the 12m array data.} The CO(2-1) emission identified to be part of a clump structure accounts for approximately 82\% of the total observed CO mass in NGC~1313 and 95\% in NGC~7793, which implies that the molecular gas in NGC~1313 is more likely to be diffuse. {The fact  that NGC~1313 has nearly twice the neutral HI gas as NGC~7793 (see Table\,\ref{tab:intro comparison}), but {about half the} molecular gas, suggests that there is an interesting difference in the balance between gas phases in these two galaxies and a thorough accounting of the total gas mass would be interesting for future work. }

\subsection{Velocity Dispersion}

We calculate the velocity dispersion for each structure by finding the intensity-weighted mean line profile and fitting a Gaussian, resulting in a fitted $\sigma_v$. We then deconvolve this $\sigma_v$ with the velocity channel width of the observations, 1.33 km s$^{-1}$, converted from FWHM to $\sigma$ with FWHM$=2.35\sigma$. The reported error in $\sigma_v$ comes from the error in the fitting method, propagated through the deconvolution. After deconvolving $\sigma_v$, measurements that were smaller than the channel width return non-number values and so are dropped from the analysis. We also remove from our analysis any structures that have a deconvolved $\sigma_v$ less than a tenth of the velocity resolution. Throughout this work, we also use the term ``linewidth'' to refer to $\sigma_v$.

\subsection{Radii}

To determine the sizes of the structures, we fit ellipses to the half-power contours and take the two axes as the HWHM of the structure. We convert these HWHM measurements to a $\sigma_R$ by approximating $\sigma_R=$HWHM$\times2/2.35$, then multiply $\sigma_R$ by 1.91 to get an ``effective radius'' \citep{Solomon87} in each axis.  We deconvolve each axis with the radius of the beam, 6.5~pc, then take the geometric mean of the axes to get a final $R$. 
The reported error in the size is determined by how non-circular the structure is, added in quadrature with a measurement error of half a pixel size, propagated through the deconvolution.
After deconvolving the radii, measurements that were smaller than the beam size return non-number values and so are dropped from the analysis. We also remove from our analysis any structures that have a reported $R$ less than a tenth of the beam size.

Of the structures that were removed because they were below the resolution limits, there were 113 clumps and 88 dendrogram leaves in NGC~1313 and 166 clumps and 116 dendrogram leaves in NGC~7793. These removed structures account for 4\% of the total clump mass in NGC~1313 and 2\% in NGC~7793. 

\subsection{Derived Quantities}

From the measured mass, radius, and linewidth above, we calculate other properties of the structures, including the surface density, $\Sigma$, the virial parameter, \alphavir,  the external pressure, $P_e$, and the free-fall time, $t_{ff}$. The surface density is simply the total mass divided by the structure's area.

The virial parameter is the ratio between twice the cloud's kinetic energy and its gravitational energy, so that a value of one indicates that the cloud is in virial equilibrium. 
{While a negative total energy balance (\alphavir<2) could be considered the threshold for a cloud's boundedness, \cite{Dib07} demonstrate that the connection between boundedness and gravitational collapse is much more complicated, and clouds with a simple negative energy balance do not always collapse. By ``collapse'', we mean the process by which parts of the cloud condenses and forms stars, while other parts of the cloud are dissipated. Furthermore, \citep{Zweibel90} show that more careful calculations of the virial parameter that take into account additional forces from external pressure and magnetic fields can alter the formula for virial equilibrium by factors of more than two. }

{Consequently, we take a more stringent threshold for gravitational collapse of \alphavir<1.  
Values greater than two suggest that the cloud is not gravitationally bound and must either disperse or be constrained by an external pressure, while values less than one suggest that the cloud is dominated by gravity and will likely begin gravitational collapse. Values of \alphavir between one and two should be interpreted cautiously.} We calculate the virial parameter for each structure with the equation 

\begin{equation}
    \alpha_\text{vir} = \frac{5 \sigma_v^2 R}{GM}.
    \label{eq:virial}
\end{equation}

{Assuming a simplified case in which the cloud is in pressure equilibrium with the surrounding medium, the external pressure will be equal to} the pressure at the edge of cloud defined by \cite{Elmegreen89} with the equation

\begin{equation}
    P_e = \frac{3\Pi M \sigma_v^2}{4\pi R^3},
    \label{eq:pressure}
\end{equation}

\noindent where $\Pi$ is defined as the ratio of the density at the edge of the cloud to the mean cloud density ($\rho_e = \Pi \langle\rho\rangle$), and here we take $\Pi=0.5$. 

The free-fall time depends only on the density of the cloud and so we calculate it with the equation

\begin{equation}
    t_{ff} = \sqrt{\frac{3\pi}{32G\rho}} = \sqrt{\frac{\pi^2 R^3}{8GM}}.
    \label{eq:tff}
\end{equation}

Tables of all of these properties and their errors, for both galaxies and for both clumps and dendrogram structures, are given in Appendix\,\ref{append: prop tables}. Given the large number of structures, we include only the first 5 entries as a demonstration of property values. The full tables are available as supplementary material.

\section{Size-Linewidth Relations} \label{sec:SL plots}

The sizes and linewidths of molecular cloud structures are expected to follow a power law relation \citep{Larson81} of the form 

\begin{equation}
    \sigma_v = a_0 R^{a_1}.
\end{equation}

The intercept ($a_0$) and slope ($a_1$) of this relation have been measured in many different environments, though the most commonly cited value is that from \cite{Solomon87}, which found $a_0 = 1.0\pm0.1$ and $a_1 = 0.5\pm0.05$ for clouds mapped in CO(1-0) in the disk of the Milky Way at 45\arcsec\ resolution, {where $\sigma_v$ is measured in km s$^{-1}$ and $R$ is measured in pc. The units of $a_0$ will vary with the value of $a_1$, but would be km s$^{-1}$ pc$^{-0.5}$ for a value of $a_1=0.5$. Since these units are dependent on $a_0$, we report the fitted values of $a_0$ without units.}

Since \cite{Solomon87}, other studies have measured the slope and intercept of this relation in a variety of environments and at many resolutions. In the Milky Way for example, \cite{Rice16} measure clouds with angular resolutions of 7.5' and measure a slope of $a_1 = 0.52\pm0.03$ and \cite{MivilleDeschenes17} measure clouds with angular resolutions of 8.5' and measure a slope of $a_1 = 0.63\pm0.3$, though neither of these studies deconvolve the sizes or linewidths with the beam and velocity channel sizes. \cite{Faesi16} measure clouds in the nearby spiral galaxy NGC~300 at 40~pc resolution with CO(2-1) and deconvolve their measurements, finding a slope of $a_1 = 0.52\pm0.2$. 
{Steeper slopes} have also been found in nearby galaxies, such as $a_1 = 0.6\pm0.1$ measured by \cite{Bolatto08} across many galaxies and with resolutions ranging from 6 to 120~pc with both CO(1-0) and CO(2-1), or $a_1 = 0.8\pm0.05$ measured by \cite{MAGMA} across the Large Magellanic Cloud (LMC) at 11~pc resolution with CO(1-0). Neither of these studies deconvolved size and linewidth measurements. Higher resolution studies (0.1-3~pc) in the LMC with CO(2-1) have performed deconvolution and found slopes ranging from $a_1 = 0.49$ to $a_1 = 0.78$ \citep{Nayak16,Indebetouw20,Wong22,Finn22}. These widely varying measurements demonstrate how resolution, molecular tracer, and the method used to measure size and linewidth all have a large effect on the measured slope. Consequently, we {focus instead on the fitted intercept of this relation, which still gives us information about the relative kinetic energy of the clouds across the full range of measured size scales. }

\begin{figure}
    \centering
    \includegraphics[width=0.45\textwidth]{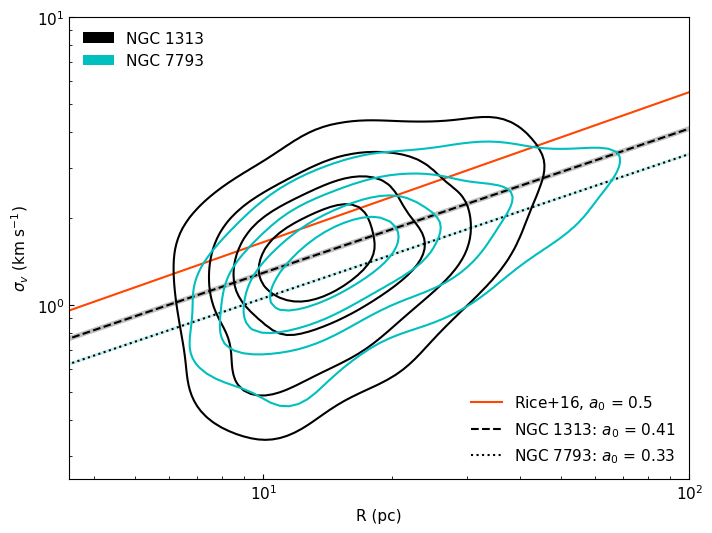}
    \caption{Kernel density estimates of the deconvolved velocity dispersions plotted against deconvolved radii of dendrogram structures in NGC~1313 and NGC~7793 with contours of 20\%, 40\%, 60\%, and 80\% of the maximum density. Also shown are the fitted power laws (where we hold the slope constant at $a_1=0.5$) with their respective $1\sigma$ errors shown as shaded regions. The resulting intercepts are printed in the bottom right corner. For comparison we show in orange the size-linewidth relation fitted in the Milky Way by \cite{Rice16}. 
    NGC~1313 has a higher intercept than NGC~7793 by more than $3\sigma$, suggesting that the molecular clouds in NGC~1313 have higher kinetic energies. }
    \label{fig:SL intercept}
\end{figure}

We show the radii and velocity dispersions of the dendrogram structures in both galaxies with a fitted power law in Figure\,\ref{fig:SL intercept}, where the power law is fit with orthogonal distance regression to take into account the error in both axes \citep[\texttt{scipy.odr; }][]{scipy}. Due to the large number of clouds, plotting all data points results in the true distribution of points being self-obscured. We instead represent their distribution with a 2-D Gaussian KDE from \texttt{scipy.stats}, using the default Scott's rule to determine the estimator bandwidth.

{We fit the intercept only of a power-law relation by holding the slope constant at a fixed value of $a_1=0.5$.}  For NGC~1313, we fit a value of $a_0=0.41\pm0.01$, and for NGC~7793 we fit a value of $a_0=0.33\pm0.01$ (Figure\,\ref{fig:SL intercept}). 
The intercept of NGC~1313 is significantly higher than that of NGC~7793 by more than $3\sigma$, which suggests that NGC~1313 has more kinetic energy in its molecular clouds than NGC~7793. 
{We also performed this fit with other values for the fixed slope in the power law to account for the wide range of fitted slopes \citep[e.g.][]{Solomon87, Bolatto08, MAGMA, Faesi16, Nayak16, MivilleDeschenes17, Indebetouw20, Wong22, Finn22}. Changing the fixed slope affects the values of the fitted intercepts, but not the conclusion that NGC~1313 has a significantly higher intercept and so more kinetic energy.}

\section{Virialization} \label{sec: virial plot}

\begin{figure}
    \centering
    \includegraphics[width=0.45\textwidth]{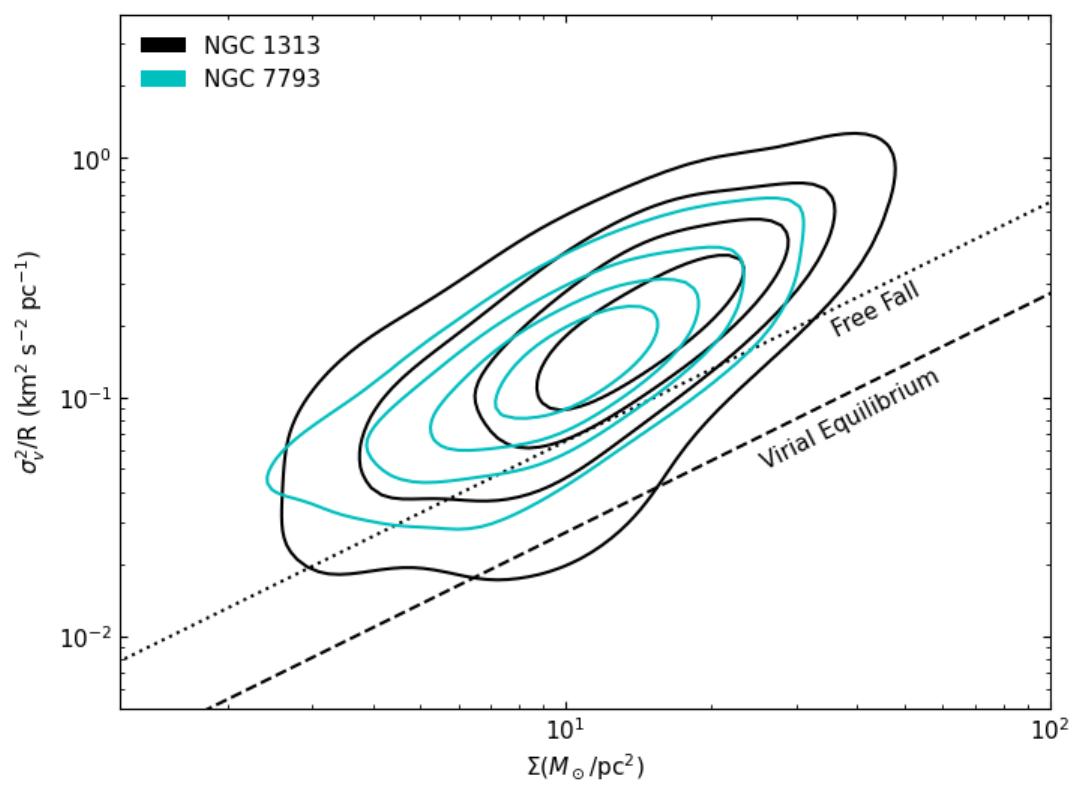}
    \caption{Kernel density estimates of the velocity metric plotted against surface density of the dendrogram structures in NGC~1313 and NGC~7793, with contours of 20\%, 40\%, 60\%, and 80\% of the maximum density. The dashed line shows where clouds in virial equilibrium would fall, above the line being dominated by kinetic energy. Clouds undergoing free-fall collapse would also have enhanced kinetic energy and fall along the dotted line. The structures in both galaxies are mostly super-virial, though the clouds in NGC~1313 appear to have more scatter. }
    \label{fig:virial}
\end{figure}

To investigate the gravitational balance of clouds in the two galaxies, we plot the velocity metric, $\sigma_v^2/R$, against the surface density, $\Sigma$, of each dendrogram structure. The results are shown in Figure\,\ref{fig:virial}, along with a line indicating where clouds in virial equilibrium (\alphavir=1) would fall in the plot. Falling above the virial equilibrium line indicates that the clouds are super-virial and dominated by kinetic energy, while clouds below the line would be sub-virial and likely to {begin forming stars}. Being super-virial, {especially when \alphavir>2}, implies that the clouds are likely to disperse due to their kinetic energy. However, the kinetic energy of clouds could also be enhanced if they already have begun free-fall collapse, in which case they would circumstantially fall along the line where \alphavir=2 in Figure\,\ref{fig:virial}. The virial equilibrium line only considers the gravitational and kinetic energies of the cloud, but external pressure or magnetic fields could also affect the boundedness of the clouds by suppressing collapse or dispersal. These environmental effects have been shown to be important in simulations of GMCs in spiral galaxies \citep{Baba17}.   

Both galaxies fall mostly above the virial equilibrium line, suggesting that they are dominated by kinetic energy. Some clouds are consistent with being in free fall, which would enhance their observed kinetic energy. Alternatively, excess kinetic energy could indicate that many of the clouds in these galaxies are not gravitationally bound, or would require an external pressure to remain bound. This is not unexpected since many studies have found a large {fraction} of unbound molecular clouds in {the Milky Way, nearby galaxy surveys, and simulations \citep{Goldsmith08,Liszt10,Dobbs11, Colombo14, MivilleDeschenes17, Sun18PHANGS, Rosolowsky21,Evans21}. }

While the two galaxies generally occupy the same parameter space in Figure\,\ref{fig:virial}, NGC~1313 appears to have a larger scatter and have more clouds close to virial equilibrium. This could suggest that more of the clouds in NGC~1313 are close to collapsing into stars and star clusters, which could drive the larger numbers of clusters in NGC~1313 despite its fewer number of clouds. The larger scatter of NGC~1313 clouds towards the unbound parameter space also matches expectations of molecular clouds responding to galaxy interactions, which cause clouds to become unbound \citep{Pettitt18, Nguyen18}.  {We investigate the boundedness of these structures more quantitatively by examining the spread in virial parameters, \alphavir, in Section\,\ref{sec:property comparison}. }

\section{Property Distribution Comparisons} \label{sec:property comparison}

\begin{figure*}
    \centering
    \includegraphics[width=0.42\textwidth]{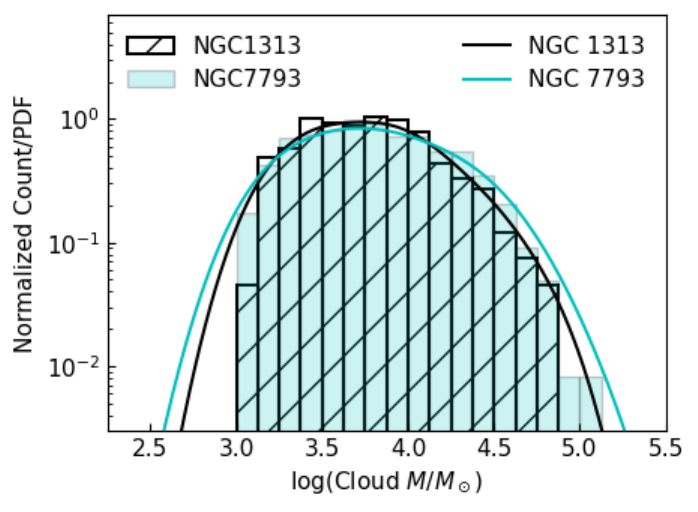}
    \includegraphics[width=0.41\textwidth]{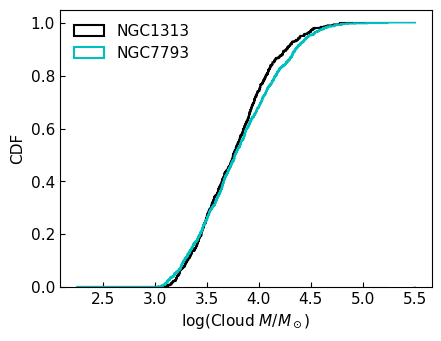}
    \includegraphics[width=0.42\textwidth]{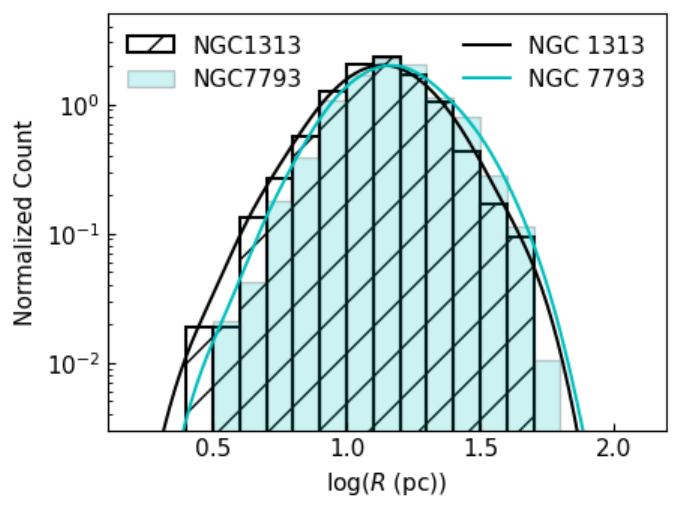}
    \includegraphics[width=0.41\textwidth]{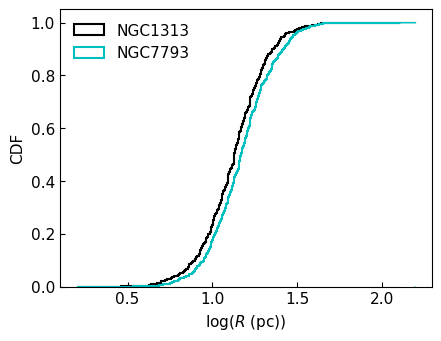}
    \includegraphics[width=0.42\textwidth]{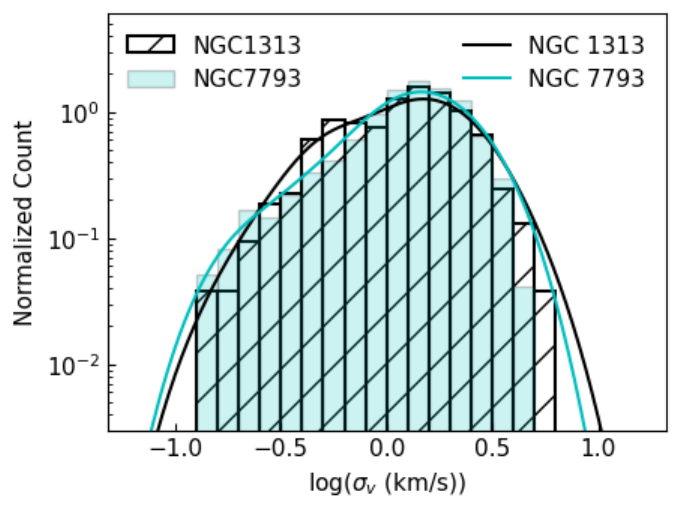}
    \includegraphics[width=0.41\textwidth]{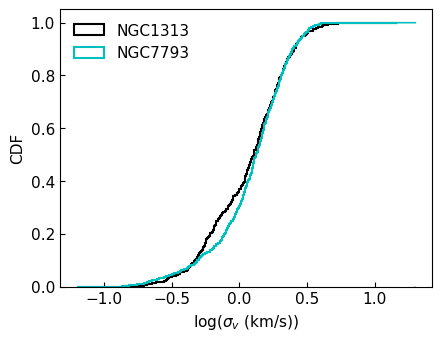}
    \caption{Distributions of the observed clump parameters for the two galaxies, using histograms and KDEs (left) and CDFs (right). We see minimal difference between the distributions except for a slightly wider mass distribution and slightly larger radii for NGC~7793.}
    \label{fig:global distributions observed}
\end{figure*}

We next investigate how the distributions of cloud properties compare between the two galaxies by looking at histograms, KDEs, and CDFs of the non-overlapping clump structures.  In Figure\,\ref{fig:global distributions observed}, we show the distributions of the three observed quantities (masses, radii, and linewidths) of the two galaxies. By eye, these distributions appear very similar, with NGC~7793 having a slightly broader distribution of masses at both the high and low mass end, and shifted to slightly larger radii.

\begin{figure*}
    \centering
    \includegraphics[width=0.42\textwidth]{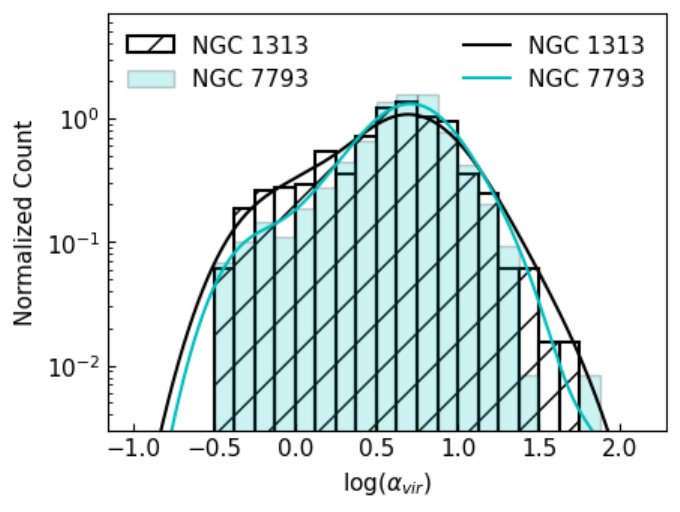}
    \includegraphics[width=0.41\textwidth]{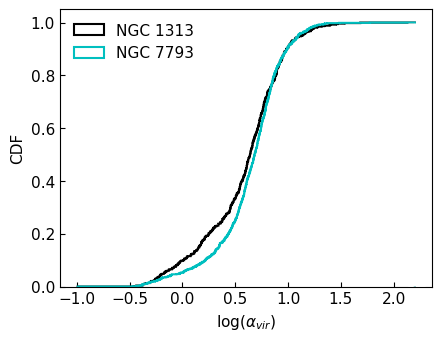} 
    \includegraphics[width=0.42\textwidth]{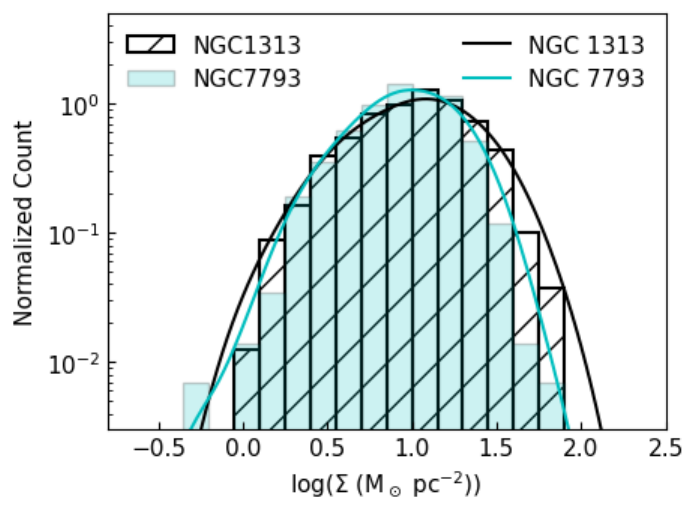}
    \includegraphics[width=0.41\textwidth]{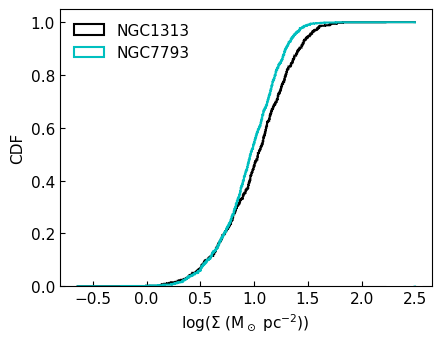}
    \includegraphics[width=0.42\textwidth]{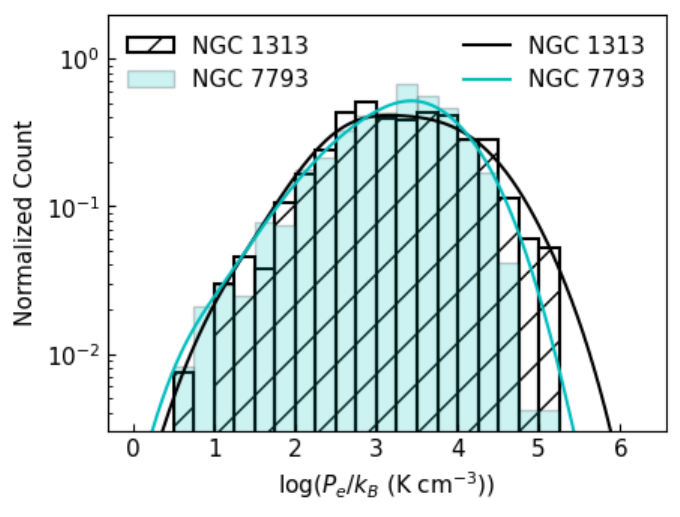}
    \includegraphics[width=0.41\textwidth]{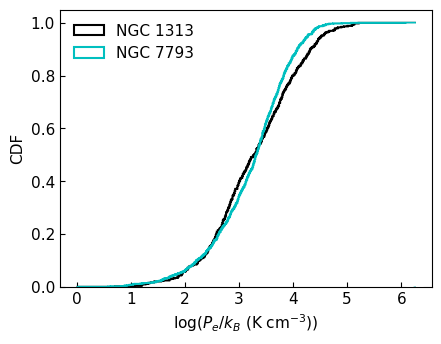}
    \includegraphics[width=0.42\textwidth]{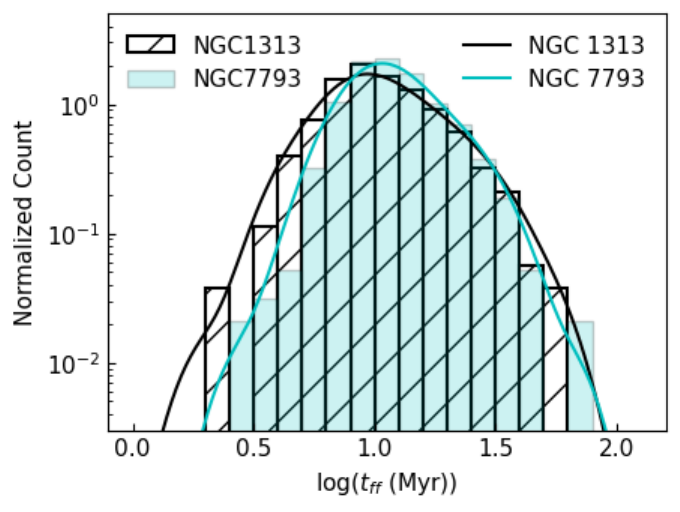}
    \includegraphics[width=0.41\textwidth]{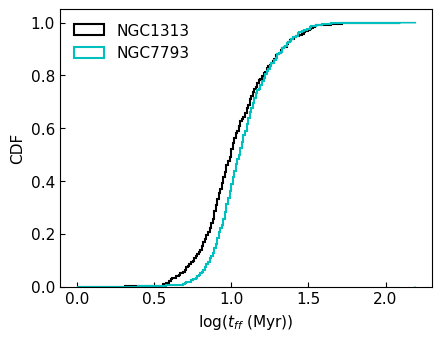}
    \caption{Distributions of the derived parameters for the two galaxies, using histograms and KDEs (left) and CDFs (right). NGC~1313 appears to have more clouds at low \alphavir, higher surface densities, higher external pressures, and lower free-fall times than NGC~7793.}
    \label{fig:global distributions derived}
\end{figure*}

We see a bit more difference between the galaxies in the properties derived from $M$, $R$, and $\sigma_v$. We show the distributions of the virial parameter, surface density, external pressure, and free-fall time in Figure\,\ref{fig:global distributions derived}. 
From these we see that NGC~1313 has more clouds at low \alphavir, which we expected from Section\,\ref{sec: virial plot}, as well as higher surface densities, higher external pressures, and lower free-fall times. 

{To better quantify the difference between property distributions, we consider the two-sample version of both the Kolmogorov-Smirnoff (KS) test and the Anderson-Darling (AD) test, where the $p$-value indicates the probability that the two samples are pulled from the same distribution. A $p$-value of less than 5\% is generally taken to be statistically significant. } 
The KS test is most sensitive to the center of the distribution, while the AD test is more sensitive to the tails. 
{However, because there are so many samples (in this case clouds), } the comparison of every property for the two galaxies have $p$-values less than 5\%, and most are also much less than 1\%. This is likely an over-representation of how different the distributions truly are though because the sample sizes of clouds and clusters for the galaxies is so large. 
As discussed in \citep{Lazariv18}, as the sample size becomes larger, KS and AD tests have increasingly higher power to discern small differences in the distributions. However, these tests do not take into account the error in the parameters, and so at large sample sizes, these tests can discern differences that are smaller than the error in the measurements, which we consider unreliable. 

To combat overpowered statistical tests, we take random subsamples of the distribution for each galaxy and perform a KS and AD test, using a sample size instead of 65. After taking 1000 random subsamples, we report the average $p$-values for each property comparison as the bootstrapped result. We show these results for each property in Table\,\ref{tab:global ks tests}. 
With this bootstrapped statistic, the only property that crosses a 5\% threshold for significance for both KS and AD tests is the distributions of cluster masses, {for both the full population and including only young clusters}. This emphasizes how the cluster populations of the two galaxies appear significantly different, but none of the cloud properties match that level of difference. 

{We recommend caution when interpreting the bootstrapped KS and AD tests, since the $p$-value of the bootstrapped test does not necessarily indicate a statistically significant difference in the two populations. 
Choosing a sufficiently large sample size causes every property to appear significantly different between the two galaxies, and similarly a sufficiently small sample size makes no property have a statistically significant result.
Rather, we use these results only to determine which properties show the biggest difference between the galaxies relative to the other properties. }

\begin{table}
    \centering
    \caption{Boostrapped two-sample KS and AD test results between NGC~1313 and NGC~7793 global properties}
    \begin{tabular}{c c c}
    \hline
    \hline
        Parameter & KS $p$-value & AD $p$-value \\
        \hline
        Cluster Mass        & 0.04    & 0.02 \\
        {Young Cluster Mass} & {0.02} & {0.01} \\
        Cluster Age         & 0.23    & 0.09 \\
        Cloud Mass          & 0.46    & 0.20  \\
        Radius              & 0.37    & 0.16  \\
        Linewidth           & 0.46    & 0.20  \\
        Virial parameter    & 0.40    & 0.18  \\
        Surface density     & 0.34    & 0.15  \\
        Pressure            & 0.40    & 0.18  \\
        Free-fall time      & 0.20    & 0.10  \\
        \hline
    \end{tabular}
    \label{tab:global ks tests}
\end{table}

These plots and tests indicate that overall, the cloud properties between the two galaxies are not particularly different. The greater number of massive clusters in NGC~1313 compared to NGC~7793 may be driven in part by higher surface densities and pressures. The lower virial parameters and free-fall times in NGC~1313 may suggest that more clouds are close to gravitational {condensation} and {are likely to form stars more quickly and with greater efficiency}, which could then result in more clusters being formed on this cloud scale. 

\begin{figure}
    \centering
    \includegraphics[width=0.5\textwidth]{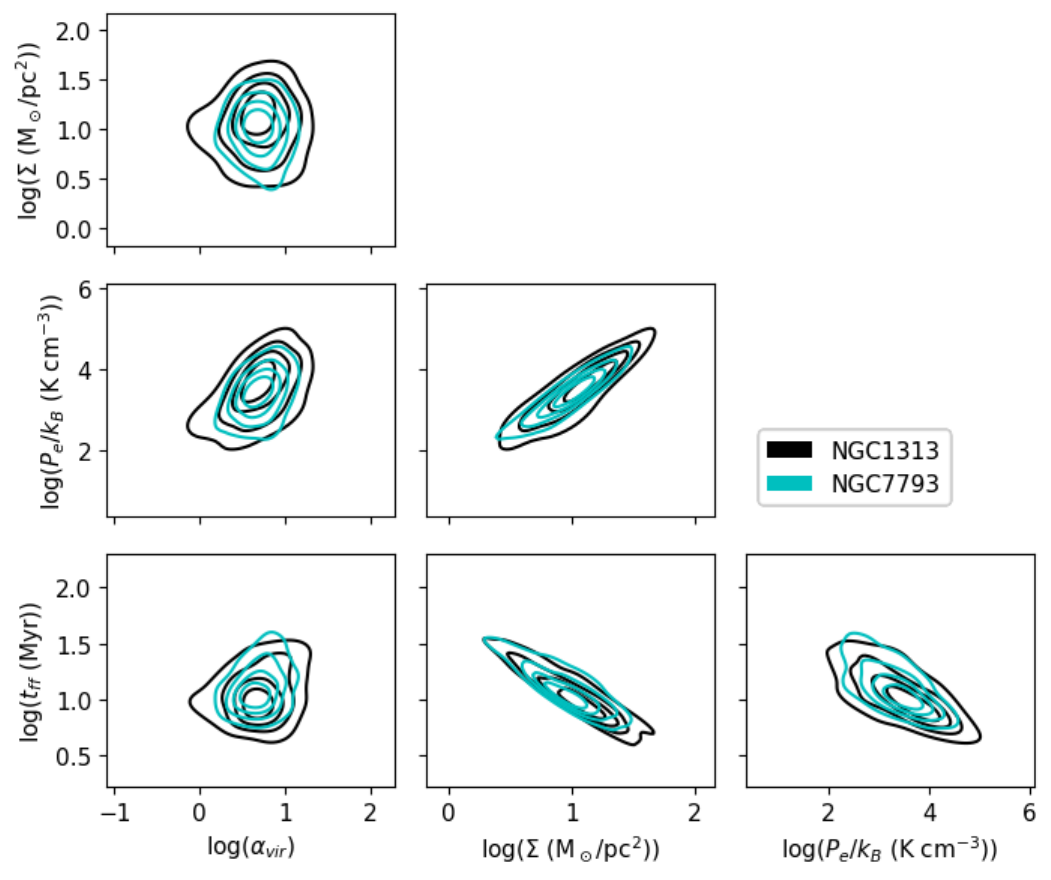}
    \caption{Gaussian KDEs of the derived dendrogram structure properties plotted against each other for both NGC~1313 and NGC~7793, with contours of 20\%, 40\%, 60\%, and 80\% of the maximum density. The virial parameters are much less correlated with the other three properties. }
    \label{fig:derived props corner}
\end{figure}

To better understand if the highest-surface density clouds in NGC~1313 represent a high-density tail of collapsing clouds, we also plot all of the derived cloud properties against one another for both galaxies in Figure\,\ref{fig:derived props corner}. These plots show that the surface density, pressure, and free-fall time are all closely correlated with one another, but the virial parameters are much less correlated with all three. This matches results from \cite{Sun22} that \alphavir shows the least correlation with other clouds properties while most others are well-correlated, especially the surface density (which is unsurprising since these parameters are calculated from the same three measurements of $M$, $R$, and $\sigma_v$). The NGC~1313 structures with low virial parameters that may be closest to collapsing do not have particularly high or low surface densities or free-fall times, and slightly low external pressures. This would suggest then that the high surface {densities} in NGC~1313 are not high because {the clouds} are collapsing, although the higher surface density clouds also have lower free-fall times. 

{
We note that the cloud masses found in this study have the unusual feature of being smaller than the maximum cluster masses seen in Figure\,\ref{fig:global cluster distributions}. For young clusters, this difference is much smaller, with the maximum young cluster masses in NGC~7793 well below the maximum cloud masses, while the young cluster masses in NGC~1313 are comparable to the maximum cloud masses. This could be an indication that the star formation efficiency is much higher in NGC~1313 than it is in NGC~7793, which would align with recent findings by \cite{Polak23} that in simulations, higher mass star clusters form with a greater efficiency than lower mass star clusters. However, even with a higher than usual formation efficiency, we do not expect to see clusters more massive than the most massive clouds. This observation could be in part caused by missing flux at the largest size scales since our observations do not include ACA 7m or TP data. It could also be an indication that cloud and cluster masses are declining over time in the galaxies. 
}

\section{Property Spatial Distributions} \label{sec:spatial distributions}

\begin{figure}
    \centering
    \includegraphics[width=0.48\textwidth]{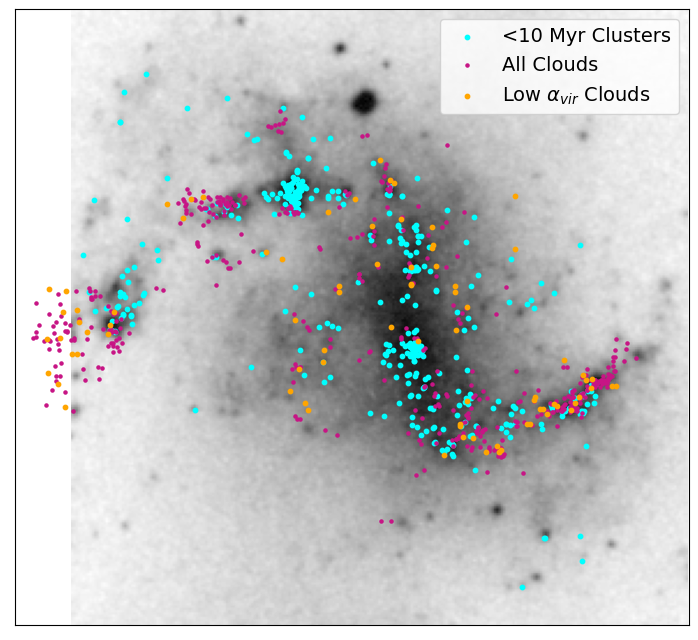}
    \includegraphics[width=0.48\textwidth]{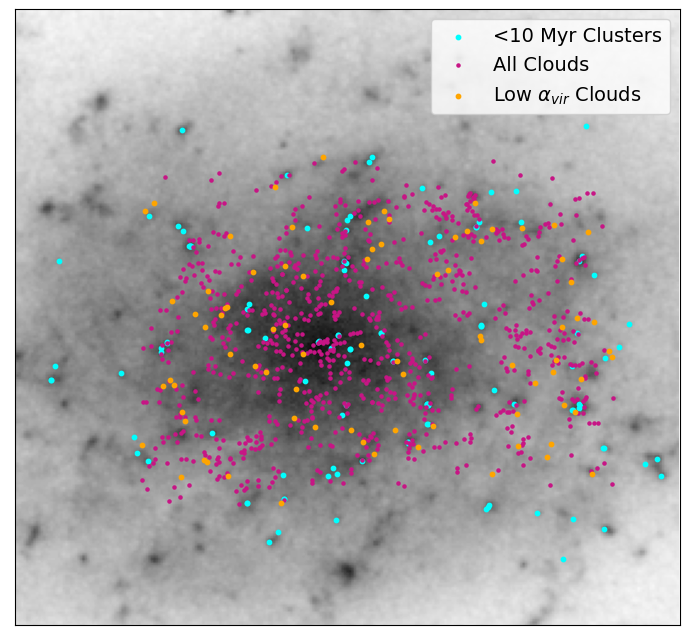}
    \caption{{Locations of the youngest clusters ($<10$ Myr; cyan), all molecular clumps (magenta), and molecular clumps with \alphavir$<1.5$ (orange) in NGC~1313 (top) and NGC~7793 (bottom). The grayscale backgrounds are the DSS2-red images. }}
    \label{fig:bound cloud maps}
\end{figure}

We next look for correlations between the clouds that are more gravitationally bound as seen in Figures\,\ref{fig:virial} and \ref{fig:global distributions derived} and the youngest of the identified clusters. In Figure\,\ref{fig:bound cloud maps}, we show for each galaxy the locations of the clouds that have \alphavir$<1.5$ and the youngest clusters ($<10$ Myr), as well as the locations of all the clouds. Here we use the clump definitions rather than the dendrogram structures for the clouds. In NGC~7793, both the clouds and the clusters have a scattered spatial distribution, as would be expected for a flocculent galaxy. In NGC~1313, the clouds and clusters are primarily found in the spiral arms and the central bar, with smaller populations in the interarm regions. 

{
To determine if the bound clouds are more closely associated with the young cluster population than the general cloud population, we create density maps of each and calculate the correlation coefficient between them. These density maps split the area of the galaxy into cells and count the number of objects (either clouds, bound clouds, or young clusters) in each cell. To mitigate the effects that our chosen cell size has on the results, we repeat this process with a variety of cell sizes ranging from 50 to 200 pixels in the ALMA maps (5\arcsec to 20\arcsec, or 110~pc to 450~pc), as well as a variety of \alphavir thresholds to determine cloud boundedness, ranging from values of 1 to 3. }

{In every case of cell size and \alphavir threshold, the bound clouds are not more correlated with young clusters than the rest of the cloud population in either galaxy. Averaging over all the cell sizes and \alphavir thresholds in NGC~1313, the correlation coefficient of the young clusters with the bound clouds is 0.24$\pm$0.06 and with the general cloud population is 0.32$\pm$0.05. In NGC~7793, the correlation coefficients are 0.36$\pm$0.18 and 0.44$\pm$0.13 for the bound clouds and all clouds, respectively. 
This suggests that the bound clouds are not more associated with young clusters than the general cloud population. NGC~7793 has slightly higher correlation coefficients than NGC~1313, though this difference is not statistically significant when we consider how the coefficients vary with different cell sizes and \alphavir thresholds. 
This conclusion is the same if we consider the full cluster population as well instead of only the youngest clusters. 
}

{We also measure the spatial correlation coefficients between the average values of the various cloud properties and the presence of young clusters, the average masses of young clusters, and the total mass of young clusters. We use the same range of cell sizes as above to make maps of these properties and compute their correlation coefficients. In both galaxies, none of the properties show a strong correlation with any of the cluster metrics. The largest coefficient averaged across cell sizes was 0.33 for the correlation between cloud surface density in NGC~1313 and the density of clusters. This further demonstrates that no individual cloud property appears uniquely correlated to the star clusters. 
}

\begin{figure}
    \centering
    \includegraphics[width=0.45\textwidth]{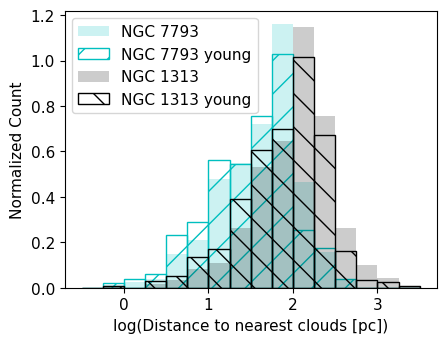}
    \caption{{Histograms of the distances between clusters and the centers of their nearest molecular cloud for both the full cluster population (solid) and the young clusters with ages <10 Myr (hatched). The young clusters appear slightly closer to molecular clouds than the general cluster population, though this is not a strong difference. There is a much larger difference in the cluster-cloud distances between the two galaxies. We hypothesize this is because the stronger spiral density waves in NGC~1313 more efficiently separate the molecular gas from the stars. }}
    \label{fig:cloud distance hist}
\end{figure}

{To further characterize the connection between the molecular clouds and the clusters, we plot histograms in Figure\,\ref{fig:cloud distance hist} of the distances between the clusters and centers of their nearest molecular clouds. We do this using the clump definitions and for both the full cluster populations as well as only the young cluster populations with ages <10 Myr, including only clusters that fall within the ALMA observational footprints. In this figure, we see that the young clusters appear slightly closer to their nearest clouds than the full cluster population for both galaxies as we would expect. This agrees with the results of \cite{Grasha18} in NGC~7793. 
This difference between the young and old clusters' distances to clouds is notably weaker than that found in M51 by \cite{Grasha19}. The weakness of this trend in NGC~1313 and NGC~7793 agrees with our finding little correlation  between the cloud properties and the presence of clusters. Including younger clusters that are closer to the generation of stars being formed by the clouds would likely result in a closer correlation.}

{Figure\,\ref{fig:cloud distance hist} shows a much larger difference in the cluster-cloud distances when comparing the two galaxies. Both the full cluster population and the young clusters of NGC~1313 are further from molecular clouds than those in NGC~7793. We believe the most likely explanation for this is the morphology of NGC~1313, where most of the molecular gas is swept up in the two large spiral arms while the star clusters get left behind, creating large separations between the clusters and clouds. The effect where star clusters get left behind by the spiral arms was demonstrated in other LEGUS galaxies by \cite{Shabani18}. This effect is much smaller in NGC~7793 where the spiral arms are weaker and closer to each other. }

\section{Discussion} \label{sec:discussion}

Overall, the measured cloud properties in the two galaxies are surprisingly similar, especially given the difference in massive star cluster formation between the two. There are slight differences in the distributions of cloud masses, sizes, and linewidths, and these small differences compound to create larger differences in their densities and energy balances. This suggests that star formation variations can be driven by relatively small shifts in these properties, at least on the size scales measured here. 

The primary differences that we find between the cloud properties in the two galaxies are that NGC~1313 has higher kinetic energies per spatial scale (Figure\,\ref{fig:SL intercept}), its clouds are closer to virial equilibrium (Figures\,\ref{fig:virial} and \ref{fig:global distributions derived}), and its clouds have higher surface densities, pressures, and shorter free-fall times (Figure\,\ref{fig:global distributions derived}). These slightly more extreme properties in NGC~1313 are likely what drive the higher rate of massive star cluster formation in the galaxy. {We also see some evidence that NGC~1313 may have higher star formation efficiency than NGC~7793 based on the relative masses of the most massive clouds in each galaxy to the most massive young clusters in each galaxy. }

Another difference in these two galaxies is in how {the different} gas phases appear to be balanced. NGC~7793 has significantly more molecular gas mass than NGC~1313 and nearly all of that molecular mass is accounted for with identified clouds, suggesting that NGC~1313 has more diffuse molecular gas than NGC~7793.  NGC~1313 also has more than twice the neutral hydrogen gas and more than twice the star formation rate as NGC~7793. This suggests that the consumption time of the gas (roughly the molecular mass divided by SFR) is much shorter in NGC~1313 than in NGC~7793. This also mirrors the lower free-fall times seen in Figure\,\ref{fig:global distributions derived}. It may be that the strong spiral density waves in NGC~1313 are important for perturbing molecular clouds to {condense and begin star formation} and then also shear these clouds apart to create greater quantities of diffuse molecular gas. Meanwhile in NGC~7793, the clouds are less likely to be perturbed and so can exist for a longer time in a dormant state without collapsing or being sheared apart.

However, it is still surprising how small the differences in cloud properties are when compared with the large difference in star formation outcomes. 
It is possible that this is because the LEGUS catalog primarily observes clusters that are optically visible and so have emerged from their natal material. It makes sense then that the cloud properties are not closely correlated with their presence as seen in Section\,\ref{sec:spatial distributions}. \cite{Messa21} searched for embedded clusters in NGC~1313 in the near infrared and found that up to 60\% of the clusters they identified are not accounted for in UV-optical catalogs. A search for embedded clusters in NGC~7793 was performed by \cite{Elmegreen19} using Spitzer data at 8~$\mu$m wavelengths. These embedded clusters would likely show a closer correlation with the molecular gas and its properties. 

For future study, we recommend undertaking identical searches for embedded sources in the two galaxies so that the younger cluster populations can be directly compared. Such a study may show that the current generation of star clusters forming in the two galaxies are more similar than previous generations seen in the LEGUS catalogs, which would explain the relatively similar cloud properties. Using young, embedded clusters would also likely reveal better insights into which cloud properties are most closely correlated with ongoing star formation. This analysis could reveal a closer correlation of young, embedded clusters with virially bound clouds, which we would expect given the tendency towards collapse and the high dependence of star formation efficiency on the boundedness of clouds \citep{Kim21,Evans22}. 

We also note that in Figures\,\ref{fig:RGB images} and \ref{fig:bound cloud maps} that map the locations and ages of the LEGUS star clusters, it is clear that in NGC~1313 especially there are differences in the level of star formation among different regions in the galaxy. We examine how the cloud properties of these two galaxies vary by sub-galactic region in a follow-up study, \cite{Finn23b}.

\section{Conclusions} \label{sec:conclusions}

We present a comparison of the molecular gas of two spiral galaxies from the LEGUS sample \citep{Calzetti15}, NGC~1313 and NGC~7793, observed in CO(2-1) with ALMA. These two galaxies have similar stellar masses, metallicities, and star formation rates, yet NGC~1313 {has formed significantly more massive star clusters in the last 10 Myr} than NGC~7793. With observations of the same molecular tracer at the same physical resolution and the same sensitivity, we compare the properties of the molecular gas to understand what differences in gas conditions give rise to such different star formation outcomes. Our major results are summarized below. 

\begin{itemize}

    \item {Comparing the clusters identified by LEGUS in the two galaxies, NGC~1313 has significantly more massive clusters ($>10^4$ M$_\odot$) and especially more young, massive clusters (<10 Myr) than NGC~7793, even after correcting for each galaxy's star formation rate and the area included by LEGUS. The mass distributions of these clusters (both the full population and only considering the young clusters) indicate that NGC~1313 is skewed towards more massive clusters than NGC~7793.}

    \item Despite having less star formation, NGC~7793 has significantly more molecular gas by mass, and more identified cloud structures. {This is in contrast to the greater amount of neutral hydrogen in NGC~1313, suggesting a strong difference in the gas phase balance in these two galaxies.} 

    \item {We fit the intercept of a size-linewidth power law relation}, holding the slope fixed at a value of $a_1=0.5$ to determine relative kinetic energies. NGC~1313 also has a significantly higher intercept than NGC~7793, suggesting that the kinetic energy of clouds in NGC~1313 are higher than in NGC~7793. 

    \item Most of the clouds in both galaxies appear to be unbound and out of virial equilibrium based on plots of their surface densities against their velocity metrics.  NGC~1313 has more clouds near virial equilibrium than NGC~7793.

    \item The distributions of cloud properties between the two galaxies show minimal differences, though the small differences in the observed masses, radii, and linewidths appear to compound into larger differences in the derived properties of virial parameter, surface density, pressure, and free-fall time. 
    
    \item NGC~1313 has lower virial parameters and free-fall times and higher surface densities and pressures than NGC~7793. The higher surface densities of NGC~1313 are not correlated with low virial parameters, which would suggest that they are not a {power-law} tail caused by clouds with enhanced surface densities because they are collapsing. 

    \item The most gravitationally bound clouds do not appear any more spatially correlated with young clusters than the general cloud population. We also find that none of the average cloud properties show a strong spatial correlation with the presence, average mass, or total mass of young star clusters. This may be because the clusters used in this study are {too old to be} associated with their natal molecular gas. Comparing the cloud properties to younger, embedded clusters in both galaxies would be an interesting focus of future work.

    \item The large difference in cluster formation between the two galaxies may be driven by perturbations from the spiral density waves in the arms of NGC~1313 causing slightly more extreme cloud properties and inducing collapse.

\end{itemize}

\begin{acknowledgements}

This material is based upon work supported by the National Science Foundation Graduate Research Fellowship Program under Grant No. 1842490. Any opinions, findings, and conclusions or recommendations expressed in this material are those of the author(s) and do not necessarily reflect the views of the National Science Foundation. 

KG is supported by the Australian Research Council through the Discovery Early Career Researcher Award (DECRA) Fellowship (project number DE220100766) funded by the Australian Government. 
KG is supported by the Australian Research Council Centre of Excellence for All Sky Astrophysics in 3 Dimensions (ASTRO~3D), through project number CE170100013. 
MRK acknowledges funding from the Australian Research Council through Laureate Fellowship LF220100020.

This paper makes use of the following ALMA data: ADS/JAO.ALMA\#2015.1.00782.S. ALMA is a partnership of ESO (representing its member states), NSF (USA) and NINS (Japan), together with NRC (Canada), NSC and ASIAA (Taiwan), and KASI (Republic of Korea), in cooperation with the Republic of Chile. The Joint ALMA Observatory is operated by ESO, AUI/NRAO and NAOJ. The National Radio Astronomy Observatory is a facility of the National Science Foundation operated under cooperative agreement by Associated Universities, Inc.

These data are associated with the HST GO Program 13364 (PI D. Calzetti). Support for this program was provided by NASA through grants from the Space Telescope Science Institute. Based on observations obtained with the NASA/ESA Hubble Space Telescope, at the Space Telescope Science Institute, which is operated by the Association of Universities or Research in Astronomy, Inc., under NASA contract NAS5-26555.

\facility{ALMA, HST (WFC3, ACS)}

\software{Pipeline-CASA51-P2-B v.40896 \citep{Davis21}, 
    CASA \citep[v.5.1.1-5, v.5.6.1; ][]{McMullin07}, 
    \texttt{astrodendro} \citep{Rosolowsky08}, 
    \texttt{quickclump} \citep{Sidorin17}, 
    Astropy \citep{astropy}, 
    Matplotlib \citep{matplotlib}, 
    NumPy \citep{numpy}, 
    SciPy \citep{scipy},
}

\end{acknowledgements}

\bibliographystyle{aasjournal}
\bibliography{references.bib}

\appendix

\section{Cloud Property Tables} \label{append: prop tables}

We include demonstrative tables of the properties derived in Section\,\ref{sec:properties}, where only the first 5 entries are shown. The full tables are available as supplementary materials with the paper. We include tables for NGC~1313 dendrogram structures (Table\,\ref{tab:ngc1313 dendro props}) and clumps (Table\,\ref{tab:ngc1313 clump props}), and NGC~7793 dendrogram structures (Table\,\ref{tab:ngc7793 dendro props}) and clumps (Table\,\ref{tab:ngc7793 clump props}).

\begin{table*}[h]
    \centering
    \caption{Catalog of NGC~1313 dendrogram properties}

\csvreader[tabular=c|c c c c c c c c c c c c,
    table head=\hline \hline ID & RA & Dec & $W_{CO}$ & CO$_\text{max}$ & Mass & Radius & $\sigma_v$ & \alphavir & $\Sigma$ & log($P_e / k_B$) & $t_{ff}$  \\ 
     & (deg) & (deg) & (K km s$^{-1}$) & (K) & ($\times10^3$ M$_\odot$) & (pc) & (km s$^{-1}$) & & (M$_\odot$ pc$^{-2}$) & (K cm$^{-3}$) & (Myr) \\
    \hline, 
    filter={\value{csvrow}<5}]
    {NGC1313ClumpProps_Table.dat}
    {num=\num, RA=\RA, Dec=\Dec, WCO21=\wco, errWCO21=\ewco, COmax=\max, massnew=\mass, errmassnew=\emass, sigvnew=\sv, errsigvnew=\esv, Rnew=\R, errRnew=\eR, alpha=\alph, alphaerr=\ealph, surfacedensity=\Sig, surfacedensityerr=\eSig, Pturb=\Pe, Pturberr=\ePe, tff=\tf, tfferr=\etf}
    {\num
    & \RA
    & \Dec
    & \wco$\pm$\ewco 
    & \max
    & \mass$\pm$\emass
    & \R$\pm$\eR
    & \sv$\pm$\esv
    & \alph$\pm$\ealph
    & \Sig$\pm$\eSig
    & \Pe$\pm$\ePe
    & \tf$\pm$\etf
    }
    \label{tab:ngc1313 dendro props}
\end{table*}

\begin{table*}[h]
    \centering
    \caption{Catalog of NGC~1313 clump properties}

\csvreader[tabular=c|c c c c c c c c c c c c,
    table head=\hline \hline ID & RA & Dec & $W_{CO}$ & CO$_\text{max}$ & Mass & Radius & $\sigma_v$ & \alphavir & $\Sigma$ & log($P_e / k_B$) & $t_{ff}$  \\ 
     & (deg) & (deg) & (K km s$^{-1}$) & (K) & ($\times10^3$ M$_\odot$) & (pc) & (km s$^{-1}$) & & (M$_\odot$ pc$^{-2}$) & (K cm$^{-3}$) & (Myr) \\
    \hline, 
    filter={\value{csvrow}<5}]
    {NGC1313ClumpProps_CL_Table.dat}
    {num=\num, RA=\RA, Dec=\Dec, WCO21=\wco, errWCO21=\ewco, COmax=\max, massnew=\mass, errmassnew=\emass, sigvnew=\sv, errsigvnew=\esv, Rnew=\R, errRnew=\eR, alpha=\alph, alphaerr=\ealph, surfacedensity=\Sig, surfacedensityerr=\eSig, Pturb=\Pe, Pturberr=\ePe, tff=\tf, tfferr=\etf}
    {\num
    & \RA
    & \Dec
    & \wco$\pm$\ewco 
    & \max
    & \mass$\pm$\emass
    & \R$\pm$\eR
    & \sv$\pm$\esv
    & \alph$\pm$\ealph
    & \Sig$\pm$\eSig
    & \Pe$\pm$\ePe
    & \tf$\pm$\etf
    }
    \label{tab:ngc1313 clump props}
\end{table*}

\begin{table*}[h]
    \centering
    \caption{Catalog of NGC~7793 dendrogram properties}

\csvreader[tabular=c|c c c c c c c c c c c c,
    table head=\hline \hline ID & RA & Dec & $W_{CO}$ & CO$_\text{max}$ & Mass & Radius & $\sigma_v$ & \alphavir & $\Sigma$ & log($P_e / k_B$) & $t_{ff}$  \\ 
     & (deg) & (deg) & (K km s$^{-1}$) & (K) & ($\times10^3$ M$_\odot$) & (pc) & (km s$^{-1}$) & & (M$_\odot$ pc$^{-2}$) & (K cm$^{-3}$) & (Myr) \\
    \hline, 
    filter={\value{csvrow}<5}]
    {NGC7793ClumpProps_Table.dat}
    {num=\num, RA=\RA, Dec=\Dec, WCO21=\wco, errWCO21=\ewco, COmax=\max, massnew=\mass, errmassnew=\emass, sigvnew=\sv, errsigvnew=\esv, Rnew=\R, errRnew=\eR, alpha=\alph, alphaerr=\ealph, surfacedensity=\Sig, surfacedensityerr=\eSig, Pturb=\Pe, Pturberr=\ePe, tff=\tf, tfferr=\etf}
    {\num
    & \RA
    & \Dec
    & \wco$\pm$\ewco 
    & \max
    & \mass$\pm$\emass
    & \R$\pm$\eR
    & \sv$\pm$\esv
    & \alph$\pm$\ealph
    & \Sig$\pm$\eSig
    & \Pe$\pm$\ePe
    & \tf$\pm$\etf
    }
    \label{tab:ngc7793 dendro props}
\end{table*}

\begin{table*}[h]
    \centering
    \caption{Catalog of NGC~7793 clump properties}

\csvreader[tabular=c|c c c c c c c c c c c c,
    table head=\hline \hline ID & RA & Dec & $W_{CO}$ & CO$_\text{max}$ & Mass & Radius & $\sigma_v$ & \alphavir & $\Sigma$ & log($P_e / k_B$) & $t_{ff}$  \\ 
     & (deg) & (deg) & (K km s$^{-1}$) & (K) & ($\times10^3$ M$_\odot$) & (pc) & (km s$^{-1}$) & & (M$_\odot$ pc$^{-2}$) & (K cm$^{-3}$) & (Myr) \\
    \hline, 
    filter={\value{csvrow}<5}]
    {NGC7793ClumpProps_CL_Table.dat}
    {num=\num, RA=\RA, Dec=\Dec, WCO21=\wco, errWCO21=\ewco, COmax=\max, massnew=\mass, errmassnew=\emass, sigvnew=\sv, errsigvnew=\esv, Rnew=\R, errRnew=\eR, alpha=\alph, alphaerr=\ealph, surfacedensity=\Sig, surfacedensityerr=\eSig, Pturb=\Pe, Pturberr=\ePe, tff=\tf, tfferr=\etf}
    {\num
    & \RA
    & \Dec
    & \wco$\pm$\ewco 
    & \max
    & \mass$\pm$\emass
    & \R$\pm$\eR
    & \sv$\pm$\esv
    & \alph$\pm$\ealph
    & \Sig$\pm$\eSig
    & \Pe$\pm$\ePe
    & \tf$\pm$\etf
    }\\
    \label{tab:ngc7793 clump props}
\end{table*}

\end{document}